\documentclass[11pt,a4paper]{article}
\pdfoutput=1

\usepackage{jheppub}
\usepackage[utf8]{inputenc}
\usepackage{enumerate}
\usepackage{multirow}
\usepackage[usenames,dvipsnames]{xcolor}

\providecommand{\be}{ \begin{equation} }
\providecommand{\ee}{\end{equation}}
\providecommand{\bea}{\begin{eqnarray}}
\providecommand{\eea}{\end{eqnarray}}
\providecommand{\nn}{\nonumber}

\providecommand{\to}{\rightarrow}

\title{Natural quark mixing and inverse seesaw\\ in a left-right model with an axion}
\author{Alex G. Dias}
\author[]{and Julio Leite}
\affiliation[]{Centro de Ci\^encias Naturais e Humanas, Universidade Federal do ABC,\\
09210-580, Santo Andr\'e-SP, Brasil}
\emailAdd{alex.dias@ufabc.edu.br}
\emailAdd{julio.leite@ufabc.edu.br}
\abstract{We consider a minimal left-right model with a Peccei-Quinn symmetry, where generalised charge conjugation plays the role of the left-right symmetry. We show how the spontaneous breaking of the Peccei-Quinn symmetry by a scalar singlet can provide us with solutions not only to the strong CP and dark matter problems but can also help to generate naturally suppressed off-diagonal CKM elements and small neutrino masses via the inverse seesaw mechanism. For this, we make use of an economical scalar sector composed of a bi-doublet, two doublets and a singlet only. As a result of the new gauge bosons and neutrinos, the neutrinoless double beta decay, as well as lepton flavour violating processes, receives new contributions which can, in principle, become relevant due to the low-scale nature of the inverse seesaw mechanism. The model can easily accommodate all the current data on fermion masses and mixing even if the left-right scale is only high enough to evade the current experimental constraints. }
\arxivnumber{1902.08235}
\begin{document}
\maketitle
\flushbottom
\section{Introduction}

A common way to tackle some of the open questions in the Standard Model (SM) is to extend its gauge structure. One of the simplest SM extensions, based on the gauge group
\begin{equation}
SU(3)_c\otimes SU(2)_L\otimes SU(2)_R \otimes U(1)_{B-L}~,
\end{equation}
is the left-right (LR) symmetric model. In its minimal versions, LR models also feature a discrete symmetry, either the generalised parity $\mathcal{P}$ or charge conjugation $\mathcal{C}$, connecting both the $SU(2)$ groups. Even though $\mathcal{P}$ is more commonly used as the LR symmetry and, in fact, it was originally chosen as such \cite{Mohapatra:1974hk,  Mohapatra:1974gc, Senjanovic:1975rk, Senjanovic:1978ev}, the case for adopting $\mathcal{C}$ instead has the advantage of being compatible with the $SO(10)$ grand unified theory. As a result of the imposition of such symmetries, three generations of right-handed neutrinos are necessarily introduced allowing, in principle, for the implementation of the (type-I) seesaw mechanism for neutrino mass generation \cite{Minkowski:1977sc, Yanagida:1979as, GellMann:1980vs, Mohapatra:1979ia}. 

In general, generating neutrino masses via the type-I seesaw mechanism requires a very high energy scale ($v_R\sim\Lambda_{GUT}$) from which the right-handed neutrinos get their masses. In the case of the LR model, the very same $v_R$ scale breaks the original symmetry down to the SM group, and searches for new gauge bosons can be used to put a lower bound on $v_R$ of only a few TeV \cite{Sirunyan:2018pom,Aaboud:2018spl}. While the type-I seesaw mechanism can be accommodated at such a ``low'' scale, the price to be paid is the introduction of unreasonably suppressed Yukawa couplings. Thus, other variants of the seesaw mechanism that can take place at the TeV scale without requiring unnaturally suppressed Yukawas, such as the inverse seesaw \cite{Mohapatra:1986bd}, become very attractive in this context. The implementation of the inverse seesaw in some of the different versions of the LR model has been explored e.g. in refs. \cite{Aranda:2009ut, Dev:2009aw, BhupalDev:2010he, Parida:2012sq, Das:2016akd, Das:2017hmg,Brdar:2018sbk}.

When it comes to the quark sector, the minimal LR model provides quarks with masses due to their coupling to a scalar bi-doublet (and its charge conjugated) that acquires a nontrivial vacuum expectation value (vev). However, similar to the SM, the mass hierarchy among the different generations and the approximately diagonal structure of the quark mixing, CKM, matrix are left unexplained. In a recent work \cite{Dev:2018pjn}, the authors propose an elegant solution within the LR framework to the latter problem to what they refer as the ``flavour alignment'' puzzle. In their model, quark mixing is small as a result of being generated at one-loop in a LR model with a Peccei-Quinn (PQ) symmetry \cite{Peccei:1977hh, Weinberg:1977ma, Wilczek:1977pj}. This happens when, in addition to the new global symmetry, new degrees of freedom are introduced in the scalar sector of the most popular LR model. With the spontaneous breaking of the PQ symmetry at a very high scale, the strong CP problem is solved and an invisible axion arises and plays the role of a dark matter candidate. Nevertheless, in order to reproduce the current neutrino data, they argue that the LR symmetry has to be broken at $50$ TeV, which is above the most stringent constraints as well as outside the reach of current experiments. An alternative solution to the flavour alignment puzzle and the strong CP problem was later proposed in ref. \cite{Kuchimanchi:2018ebf}, where the PQ symmetry is broken softly in such a way that both the off-diagonal CKM elements and the strong CP phase are naturally suppressed for they are radiatively generated. In both ref. \cite{Dev:2018pjn} and ref. \cite{Kuchimanchi:2018ebf}, $\mathcal{P}$ is the LR symmetry.

In this paper, we face the above-mentioned problems by proposing a minimal LR model with a PQ symmetry, where $\mathcal{C}$ plays the role of the LR symmetry. In such a framework, we show that, when the global PQ symmetry is broken at a very high scale by a scalar singlet, the model provides solutions not only to the strong CP and dark matter problems but can also help to generate naturally suppressed off-diagonal elements for the quark mixing matrix and small neutrino masses via the inverse seesaw mechanism. The adopted scalar sector is compact. In addition to the PQ symmetry breaking scalar singlet $\sigma$, a $SU(2)_R$ doublet, $\eta_R$, is required to break the LR symmetry down to the SM group. A $SU(2)_L$ doublet, $\eta_L$, is also introduced to preserve the LR symmetry. Finally, to perform the last step of the symmetry breaking process, a scalar bi-doublet, $\Phi$, gets a vev at the electroweak scale. As the $SU(2)_R$ symmetry is broken by $v_R/\sqrt{2}$, the vev of $\eta_R$, the non-SM gauge bosons get masses at this scale, while the remaining vector bosons are identified with the SM ones. The usual LR fermion sector with $SU(2)_{L(R)}$ doublets which contains the SM fermions and three right-handed neutrinos is enlarged by three neutral fermion singlets required to implement the inverse seesaw. When only renormalisable Yukawa terms are assumed, the quark mixing matrix becomes trivial and some of the neutrinos remain massless. Upon the introduction of non-renormalisable operators involving the scalar $\sigma$, the small off-diagonal elements of the CKM matrix and neutrino masses are naturally obtained. Because of the new gauge bosons and the new neutrinos, all of which get masses proportional to $v_R$, the neutrinoless double beta decay ($0\nu2\beta$) receives several non-standard contributions which can, in principle, become sizeable when $v_R$ is around the TeV scale due to the low-scale nature of the inverse seesaw. Similarly, lepton flavour violating processes, such as the $\mu \to e \gamma$ decay, can receive large contributions in this framework. The model can easily accommodate all the current data on fermion masses and mixing even if the LR scale is only high enough to evade the current experimental constraints.

This work is written following the structure below. In Sec.~\ref{sec.2}, we present our model, introducing all the relevant fields and showing how they transform under the symmetries considered. The scalar potential is then shown in Sec.~\ref{sec.3}, followed by the derivation of the respective particle spectrum. Sec.~\ref{sec.4} deals with the gauge sector and its spectrum. In Sec.~\ref{sec.5}, fermion masses and mixing are studied in detail. For the quarks, we show how the structure of the CKM matrix is naturally generated in our model. While for the leptons, we show how the active neutrinos get sub-eV masses via the inverse seesaw mechanism. Section~\ref{sec.6} is devoted to the study of the most relevant contributions to $0\nu2\beta$. In section~\ref{sec.7}, we investigate how lepton flavour violating processes and heavy neutrino interactions can constraint our model. Finally, our conclusions are summarised in Sec.~\ref{sec.8}, and particular solutions for the quark and lepton sectors are displayed in the Appendices \ref{AppB} and \ref{AppC}, respectively.

\section{Symmetries and the field content}\label{sec.2}

The continuous symmetry group of our LR model can be written as
\be \label{sym}
SU(3)_c \otimes SU(2)_L \otimes SU(2)_R \otimes U(1)_{B-L}(\otimes\, U(1)_{PQ})~,
\ee
where $U(1)_{PQ}$, the global Peccei-Quinn symmetry, is in parentheses to emphasise that it is not a gauged symmetry as the others. Additionally, as already mentioned, the model is invariant under a discrete $\mathcal{C}$ symmetry. In what follows, we present all the fields in our model and define how they transform under these symmetries.

The left-handed fermions transform as $SU(2)_L$ doublets, in the same way as in the SM. Contrary to the SM though, the right-handed fermions do not come as singlets of the weak gauge group but as $SU(2)_R$ copies of the $SU(2)_L$ doublets, that is
\begin{eqnarray}\label{doublets}
L_{\alpha L} &=& \begin{pmatrix}
\nu_{\alpha L} \\ l_{\alpha L} 
\end{pmatrix}  \sim ({\bf 1}, {\bf 2}, {\bf 1}, -1)_{0}~,~~~L_{\alpha R} = \begin{pmatrix}
N_{\alpha R} \\ l_{\alpha R} 
\end{pmatrix}  \sim ({\bf 1}, {\bf 1}, {\bf 2}, -1)_{-\frac{1}{2}}~,\\
Q_{\alpha L} &=& \begin{pmatrix}
u_{\alpha L} \\ d_{\alpha L} 
\end{pmatrix}  \sim ({\bf 3}, {\bf 2}, {\bf 1}, \frac{1}{3})_{0}~,~~~Q_{\alpha R} = \begin{pmatrix}
u_{\alpha R} \\ d_{\alpha R} 
\end{pmatrix}  \sim ({\bf 3}, {\bf 1},  {\bf 2}, \frac{1}{3})_{-\frac{1}{2}}~,\nn
\end{eqnarray}
with $\alpha = 1, 2, 3$; the numbers in parentheses show how the fields transform under the $SU(3)_c$, $SU(2)_L$, $SU(2)_R$ and $U(1)_{B-L}$ gauge symmetries, respectively, while the index tells us the field transformation under the global $U(1)_{PQ}$. In addition to the fermion doublets, in order to implement the inverse seesaw, we also introduce three neutral fermion singlets,\footnote{Note that the fermion singlets $S_{\alpha R}$ do not carry lepton number.}
\be 
S_{\alpha R} \sim ({\bf 1}, {\bf 1}, {\bf 1}, 0)_{\frac{n+1}{2}}~,
\ee 
with $n$ an integer.
As shown in ref.~\cite{Abada:2014vea}, when three left-handed and three right-handed neutrinos, $\nu_{\alpha L}$ and $\nu_{\alpha R}$, are present and neutrino masses are generated via the inverse seesaw mechanism, three is also the minimum number of fermion singlets $S_{\alpha R}$ required to reproduce the current neutrino data. At last, under the discrete $\mathcal{C}$ symmetry, the fermions transform according to\footnote{Note that the singlets in our model transform trivially under this generalised charge conjugation symmetry.}
\be\label{cconjf}
\mathcal{C}:~~~~\{ L_{\alpha L}\,, \,Q_{\alpha L}\,,\, S_{\alpha R}\}~~\leftrightarrow~~ \{ (L_{\alpha R})^c\,,\, (Q_{\alpha R})^c\,,\,S_{\alpha R}\}~,
\ee 
where the $c$ on the right-hand side represents the usual charge conjugation transformation.

In the extended electroweak sector there exists a total of seven vector fields: $W_{L\mu}^i$, $W_{R\mu}^i$ and $B_\mu$, with $i = 1,2,3$, associated with the gauge groups $SU(2)_L$, $SU(2)_R$ and $U(1)_{B-L}$, respectively. These fields can be grouped into the covariant derivative acting on field multiplets. For example, the electroweak covariant derivative acting on $SU(2)_{L(R)}$ doublets, $\psi_{L(R)}$, can be defined as
\be\label{CovDer}
D_\mu \left(\psi_{L(R)}\right) \equiv \left(\partial_\mu  - i \frac{g}{2} W^j_{L(R)\mu} \tau_j-ig_{B-L} \frac{B-L}{2} B_\mu\right) \psi_{L(R)} = (\partial_\mu  - i P^{L(R)}_\mu) \psi_{L(R)} 
\ee
where $\tau_j$ are the Pauli matrices, $g_{L}=g_R=g$ as a result of the LR symmetry, and $P^{L(R)}_\mu$ can be written in a $2\times2$ form as
\be 
P^{L(R)}_\mu \equiv\frac{g}{2}\begin{pmatrix}
W_{L(R)\mu}^3 + t (B-L) B_\mu & \sqrt{2}\tilde{W}_{L(R)\mu}^+  \\ \sqrt{2}\tilde{W}_{L(R)\mu}^-  & -W_{L(R)\mu}^3 + t (B-L) B_\mu 
\end{pmatrix},
\ee
with $\sqrt{2}\tilde{W}_{L(R)\mu}^\pm = (W_{L(R)\mu}^1 \mp i W_{L(R)\mu}^2) $ and $t = g_{B-L}/g$. 

The scalar sector is compact and contains the following singlet, $SU(2)_L$ and $SU(2)_R$ doublets and bi-doublet
\bea
\sigma &\sim& ({\bf 1}, {\bf 1}, {\bf 1}, 0)_1~,\\
\eta_{L} &=& \begin{pmatrix}
\eta^0 \\ \eta^- 
\end{pmatrix}_{L}\sim ({\bf 1}, {\bf 2}, {\bf 1} , -	1)_{-\frac{n+1}{2}}~,~~
\eta_{R} = \begin{pmatrix}
\eta^0 \\ \eta^- 
\end{pmatrix}_{R}\sim ({\bf 1}, {\bf 1}, {\bf 2}, -1)_{\frac{n}{2}}~\nn\\
\Phi  &=& \begin{pmatrix}
\phi_1^0 & \phi_2^+ \\ \phi_1^- & \phi_2^0
\end{pmatrix}~  \sim ({\bf 1}, {\bf 2}, {\bf 2}, 0)_{\frac{1}{2}}~,\nn
\eea
which transform under $\mathcal{C}$ as 
\be\label{cconjs}
\mathcal{C}:~~~~~\{ \sigma\,,\, \Phi\,,\, \eta_L\} ~~\leftrightarrow~~ \{\sigma\,,\, \Phi^T\,,\, \eta_R^*\}~.
\ee
The Higgs mechanism takes place when the neutral components of such fields acquire non-trivial vevs,  breaking spontaneously the initial symmetry in eq. (\ref{sym}) down to $SU(3)_c\otimes U(1)_Q$ in three steps. In the first step $\langle \sigma \rangle = v_\sigma/\sqrt{2}$ breaks the PQ symmetry. The second stage, i.e.
\begin{eqnarray}
SU(3)_c \otimes SU(2)_L \otimes SU(2)_R \otimes U(1)_{B-L}~ \to SU(3)_c \otimes SU(2)_L\otimes U(1)_Y~,
\end{eqnarray}
is achieved when $\eta_R^0$ acquires a vev: $v_R/\sqrt{2}$. Finally, the last step of the breaking process,
\begin{eqnarray}
SU(3)_c \otimes SU(2)_L \otimes U(1)_{Y} \to SU(3)_c \otimes U(1)_{Q}~,
\end{eqnarray}
is performed by the vevs of $\Phi$:  $\langle \phi_1^0\rangle = v_{1}/\sqrt{2}$ and $\langle \phi_2^0\rangle = v_{2}/\sqrt{2}$. We assume the following hierarchy among the different energy scales $v_\sigma \gg v_R \gg v_{1} \gg v_2$. 

In summary, in Table \ref{t1} we show the $U(1)$ charges of the fermion and scalar fields, from which we can see that $U(1)_{B-L}$ is a gauged combination of $U(1)_B$ and $U(1)_L$. It is also worth mentioning that, along the lines of refs. \cite{Dias:2014osa} and \cite{Dias:2018ddy}, discrete symmetries can be used to protected our axion solution from potentially dangerous $U(1)_{PQ}$ violating gravitational corrections.

\begin{table}[t]
    \begin{center}
		\begin{tabular}{ |c|c|c|c|c|c|c|c|c|c| } 
			\hline
            {\bf Fields}  & ~$\sigma$ ~  & $Q_{\alpha L}$ & $L_{\alpha L}$ & $S_{\alpha R}$ & ~$\Phi$~ & $Q_{\alpha R}$ & $L_{\alpha R}$ & $\eta_L$  & $\eta_R$    \\\hline 
			  $U(1)_B $ & $ 0 $ & $ \frac{1}{3} $ & $ 0 $ & $ 0 $  & $ 0 $ & $ \frac{1}{3} $ & $ 0 $ &  $ 0  $ & $ 0 $ \\  \hline
              $U(1)_L $ & $ 0 $ & $ 0 $ & $ 1 $ & $ 0 $  & $ 0 $ & $ 0 $ & $ 1 $ &  $ 1  $ & $ 1  $ \\  \hline
              $U(1)_{PQ} $ & $ 1 $ & $ 0 $ & $ 0 $ & $ \frac{n+1}{2} $  & $ \frac{1}{2} $ & $ -\frac{1}{2} $ & $ -\frac{1}{2} $ &  $ -\frac{n+1}{2}  $ & $ \frac{n}{2}  $ \\  \hline
		\end{tabular}
		\caption{Field transformations under the global symmetries, where $n$ is an integer.}\label{t1}
	\end{center}
\end{table}

\section{The scalar sector}\label{sec.3}

With the scalar content and symmetries described so far, we can write down the most general renormalisable scalar potential as 
\bea\label{Vh} 
V_h &=&  \mu^2_\sigma (\sigma^*\sigma) + \mu^2(\eta_L^\dagger \eta_L+\eta_R^\dagger \eta_R) + \mu_\Phi^2 Tr(\Phi^\dagger \Phi)
+\lambda_\sigma (\sigma^*\sigma)^2 + \lambda[(\eta_L^\dagger \eta_L)^2+(\eta_R^\dagger \eta_R)^2]\nn\\
&&+
\lambda_{LR}(\eta_L^\dagger \eta_L)(\eta_R^\dagger \eta_R) + \lambda_{\Phi1} Tr(\Phi^\dagger \Phi \Phi^\dagger \Phi) +  \lambda_{\Phi2}[Tr(\Phi^\dagger \Phi)]^2 + \lambda_{\Phi 3} Tr(\Phi^\dagger \Phi\tilde{\Phi}^\dagger \tilde{\Phi})\nn \\
&&+ \frac{\lambda_{\Phi 4}}{2} Tr(\Phi^\dagger \tilde{\Phi})Tr(\tilde{\Phi}^\dagger \Phi) +\lambda_{\Phi 5} Tr(\Phi^\dagger \tilde{\Phi}\tilde{\Phi}^\dagger \Phi)  + \lambda_{\sigma\eta} (\sigma^*\sigma)( \eta_L^\dagger \eta_L+\eta_R^\dagger \eta_R ) \nn\\
&&+ \lambda_{\sigma\Phi} (\sigma^*\sigma) Tr(\Phi^\dagger \Phi)+ \lambda_{\eta\Phi} ( \eta_L^\dagger \eta_L+\eta_R^\dagger \eta_R )Tr(\Phi^\dagger \Phi)~\nn\\
&&+ \lambda_{\eta\Phi1}( \eta_L^\dagger \Phi \Phi^\dagger \eta_L+\eta_R^\dagger \Phi^\dagger \Phi\eta_R )+ \lambda_{\eta\Phi2} ( \eta_L^\dagger \tilde{\Phi} \tilde{\Phi}^\dagger \eta_L+\eta_R^\dagger \tilde{\Phi}^\dagger \tilde{\Phi}\eta_R )~,
\eea
plus a non-Hermitian term
\be\label{Vnh}
V_{nh} = \frac{f}{\sqrt{2}}\,\sigma Tr( \Phi^\dagger \tilde{\Phi} ) + h.c.
\ee
where $f$ is a dimensionful coupling and $\tilde{\Phi} = \tau_2 \Phi^*\tau_2$. In order to find the scalar spectrum, we replace the field decompositions below in the scalar potential
\bea\label{scdec}
\sigma &=& \frac{1}{\sqrt{2}} (v_\sigma + S_\sigma + i A_\sigma)~, ~~\eta_{L(R)} = \begin{pmatrix}
\frac{1}{\sqrt{2}}  (v_{L(R)}+S_{L(R)} + i A_{L(R)})\\ \eta^-_{L(R)} 
\end{pmatrix}~, \\
\Phi  &=& \begin{pmatrix}
\frac{1}{\sqrt{2}}  (v_1+S_{1} + i A_1) & \phi_2^+ \\ \phi_1^- & \frac{1}{\sqrt{2}}  (v_2+S_2 + i A_2) 
\end{pmatrix}~.\nn
\eea
From the first derivative of the potential, the constraints below follow
\begin{eqnarray} 
&& v_\sigma [2 v_\sigma^2 \lambda_\sigma + (v_L^2 + 
       v_R^2) \lambda_{\sigma\eta} + (v_1^2 + 
       v_2^2) \lambda_{\sigma\Phi} + 2 \mu_\sigma^2]=-2 f v_1 v_2~, \\
 &&       v_L [2 v_L^2 \lambda + v_R^2 \lambda_{LR} + 
   v_1^2 (\lambda_{\eta\Phi} + \
\lambda_{\eta\Phi1}) + 
   v_2^2 (\lambda_{\eta\Phi} + \
\lambda_{\eta\Phi2}) + 
   v_\sigma^2 \lambda_{\sigma\eta} + 2 \mu^2]= 0~,\nn\\
 &&  v_R [2 v_R^2 \lambda + v_L^2 \lambda_{LR} + 
   v_1^2 (\lambda_{\eta\Phi} + \
\lambda_{\eta\Phi1}) + 
   v_2^2 (\lambda_{\eta\Phi} + \
\lambda_{\eta\Phi2}) + 
   v_\sigma^2 \lambda_{\sigma\eta} + 2 \mu^2]= 0~,\nn\\
&& v_1 \left[(v_L^2 + 
       v_R^2) (\lambda_{\eta\Phi} + \lambda_{\eta
\Phi1}) + v_\sigma^2 \lambda_{\sigma\Phi} + 
    2 \left(v_1^2  \sum_{i=1}^2\lambda_{\Phi i} + 
       v_2^2 \sum_{j=2}^5\lambda_{\Phi j} + \
\mu_\Phi^2\right)\right]= -2 f v_2 v_\sigma~,\nn\\
&&  v_2 \left[(v_L^2 + 
        v_R^2) (\lambda_{\eta\Phi} + \lambda_{\eta
\Phi2}) + v_\sigma^2 \lambda_{\sigma\Phi} + 
     2 \left(v_2^2 \sum_{i=1}^2\lambda_{\Phi i} + 
        v_1^2 \sum_{j=2}^5\lambda_{\Phi j} + \
\mu_{\Phi}^2\right)\right]= -2 f v_1 v_\sigma~.\nn
\end{eqnarray}

A possible solution for the system of equations above is the asymmetric one with  $v_L=0$ and $v_R\neq0$. In this case, we also have 
\begin{equation}\label{f}
f = \frac{v_1 v_2 [v_R^2 (\lambda_{\eta\Phi1} -  \lambda_{\eta\Phi2}) + 
   2 (v_1^2-v_2^2)( \lambda_{\Phi1}-\lambda_{\Phi3}-\lambda_{\Phi4}-\lambda_{\Phi5})]}{2 (v_1^2 - v_2^2) v_\sigma}~.
\end{equation}
Similarly, it allows us to rewrite the parameters $\mu_\sigma^2,\, \mu^2$ and $\mu_\Phi^2$ in terms of the vevs and the dimensionless couplings. From eq. (\ref{f}), assuming the dimensionless couplings to be of order one, and the vevs: $v_2 = \mathcal{O}(1)$ GeV, $v_1 = \mathcal{O}(10^2)$ GeV, $v_R= \mathcal{O}(10^4)$ GeV and $v_\sigma = \mathcal{O}(10^{11})$ GeV,  we expected that $f = \mathcal{O} (10^{-4})$ GeV. The scale of the vev $v_\sigma$ is taken here to have a value within the interval where the axion -- the pseudo Nambu-Goldstone  boson of the PQ symmetry breakdown -- can be a cold dark matter candidate~\cite{Tanabashi:2018oca} (see below). 

At this point it is worth stressing that without the PQ symmetry, the scalar potential would contain terms such as $\eta_L^\dagger \Phi \eta_R$, resulting in an effective vev for $\eta_L$. As we shall see in the next sections, $v_L = 0$ (or at least $v_L$ negligible when compared to the other vevs) is paramount for the inverse seesaw mechanism to take place. Therefore, the PQ symmetry is intrinsically associated with the neutrino mass generation mechanism in our model. Previous studies involving the inverse seesaw in LR models, such as in refs.~\cite{Aranda:2009ut, Brdar:2018sbk}, have dealt with this issue differently. While in ref.~\cite{Aranda:2009ut}, the authors avoid such an issue by assuming an asymmetric LR model in which $\eta_L$ is absent, in ref.~\cite{Brdar:2018sbk} the authors argue that by keeping the dimensionful parameter $f$ small ($f\lesssim 100$ keV for $v_R\simeq 100$ TeV), the effective vev $\langle \eta_L^0\rangle$ will be small enough to make sure that the inverse seesaw is realised.

With this particular solution that minimises the potential, we can finally derive the scalar particle spectrum. The scalar sector initially contains eighteen degrees of freedom. After symmetry breaking, six of those are expected to be absorbed by the gauge sector and make the vector bosons massive. Therefore, we are left with twelve physical scalar degrees of freedom. For simplicity, we present here only the mass states and eigenvalues, while all the squared mass matrices are shown in the Appendix \ref{AppA}. 

While the following charged fields (4 degrees of freedom) are absorbed by the charged gauge bosons $W_L^\pm$ and $W_R^\pm$,
\begin{eqnarray}
G_1^\pm &=& \frac{1}{\sqrt{v_1^2+v_2^2}}[v_1\phi_1^\pm-v_2\phi_2^\pm]~,\\
G_2^\pm &=& \frac{1}{\sqrt{(v_1^2+v_2^2)[(v_1^2+v_2^2)v_R^2+(v_1^2-v_2^2)^2]}}\left\{v_R(v_1^2+v_2^2)\eta_R^\pm+(v_1^2-v_2^2)[v_2\phi_1^\pm+v_1\phi_2^\pm]\right\}~,\nn
\end{eqnarray}
the remaining charged scalar fields become massive
\begin{eqnarray}
H_1^\pm &=& \frac{1}{\sqrt{(v_1^2 + v_2^2) v_R^2 + (v_1^2 - v_2^2)^2} }\left[(v_2^2-v_1^2 )\eta_R^\pm + (v_2v_R)\phi_1^\pm  + (v_1 v_R)\phi_2^\pm\right]~,\\
H_2^\pm &=& \eta_L^\pm~,\nn
\end{eqnarray}
with the following squared masses
\begin{eqnarray}
m_{H_1^\pm}^2 &=&  \frac{(v_1^2 - 
      v_2^2)^2 + (v_1^2 + 
       v_2^2) v_R^2 }{
 2 (v_1^2 - v_2^2)}(\lambda_{\eta\Phi 2} - \lambda_{\eta\Phi 1})~, \\
m_{H_2^\pm}^2 &=& \frac{1}{2} [v_R^2 ( \lambda_{LR}-2 \lambda) + (v_1^2 - 
      v_2^2)  (\lambda_{\eta\Phi 2} - \lambda_{\eta\Phi 1})]~,\nn
\end{eqnarray}
and $(\lambda_{\eta\Phi 2}-\lambda_{\eta\Phi1})>0$, $(\lambda_{LR}-2 \lambda)>0$.

The (complex) neutral field $\eta_L^0$ does not mix with the other real fields, and gets the following mass term
\begin{equation}
m_{\eta_L^0}^2 = \frac{v_R^2}{2}(\lambda_{LR}-2\lambda)~.
\end{equation}

We consider now the CP-odd fields $A_R, A_\sigma, A_1$ and $A_2$. The first one is a Goldstone boson, $G_1 =A_R$ , which is absorbed by the neutral gauge boson $Z_R$, as defined in the next section. The other three are mixed, and after diagonalisation, we find 
\bea\label{Afields}
G_2 &=& \frac{1}{\sqrt{(v_\sigma^2+v_2^2)[v_\sigma^2(v_1^2+v_2^2)+v_1^2 v_2^2]}}[(v_\sigma v_2^2) A_\sigma + v_1(v_2^2+v_\sigma^2) A_1 - (v_2 v_\sigma^2) A_2] ,\\
A &=& \frac{1}{\sqrt{v_\sigma^2(v_1^2+v_2^2)+v_1^2 v_2^2}}[-(v_1 v_2) A_\sigma+ (v_2 v_\sigma) A_1 + (v_1 v_\sigma) A_2]~,\nn\\
a &=& \frac{1}{\sqrt{v_\sigma^2+v_2^2}}[v_\sigma A_\sigma + v_2 A_2]~,\nn
\eea
where $G_2$ is absorbed by the gauge sector, making the neutral gauge boson $Z_L$, defined in the next section, massive; $A$ is a CP-odd scalar with mass
\bea
m_A^2 &=& \frac{[v_\sigma^2(v_1^2+v_2^2)+v_1^2 v_2^2][v_R^2(\lambda_{\eta\Phi2}-\lambda_{\eta\Phi1})+2(v_1^2-v_2^2)(\lambda_{\Phi3}+\lambda_{\Phi4}+\lambda_{\Phi5}-\lambda_{\Phi1})]}{2(v_1^2-v_2^2)v_\sigma^2}\nn\\
&\simeq&  \frac{(v_1^2 + v_2^2)v_R^2}{2(v_1^2-v_2^2)}(\lambda_{\eta\Phi 2}-\lambda_{\eta\Phi 1})~,
\eea
and $a$ is the axion field that gets its mass from non-perturbative effects
\be \label{maxion}
m_a \simeq 5.7 \times \left(\frac{10^9~\text{GeV}}{f_a}\right)\text{meV}~ 
\ee
with the axion decay constant $f_a \simeq v_\sigma/N_{DW}$, where the domain wall number, in the present case, is $N_{DM}=3$. For example, taking $v_\sigma = 10^{11-10}$ GeV we can have $m_a\simeq (0.17-1.7)$ meV. Studies on the DFSZ ($N_{DM}=6$) and KSVZ ($N_{DM}=1$) models have pointed out that axions with masses around the meV and sub-meV scales, respectively, could be the dominant component of dark matter~\cite{Ringwald:2015dsf} (see also~\cite{Tanabashi:2018oca}). Although our model might be viewed as a hybrid DFSZ-KSVZ axion model since both the SM and the new heavy quarks carry $U(1)_{PQ}$ charges, we expect that the axion here with mass in the meV or sub-meV scale could act as the dominant dark matter component in the Universe. A detailed study devoted to determine the preferred mass range in which the axion in our model can play the role of the dominant dark matter component would certainly be interesting, but out of the scope of the present work.

Finally, we take a look at the CP-even fields $S_\sigma,\, S_1,\, S_2$ and $S_R$. These four real fields mix among themselves according to a $4\times4$ symmetric mass matrix. Making use of the fact that $v_\sigma$ is much larger than any other vev in the model, we consider the limit where $S_\sigma$ decouples from the other real fields. In such a limit, $S_\sigma$ gets the following mass
\begin{eqnarray}
m_{S_\sigma}^2 \simeq 2 \lambda_\sigma v_\sigma^2~.
\end{eqnarray}
Thus, the original $4\times4$ becomes a $3\times3$ matrix. To make its diagonalisation process simpler, we take the smallest vev, $v_2$, to be zero. Upon this simplification, we find the following eigenmasses
\begin{eqnarray}
m_{S_1}^2 &\simeq& v_1^2 \frac{[4 \lambda(\lambda_{\Phi 1}+\lambda_{\Phi 2})-(\lambda_{\eta\Phi }+\lambda_{\eta\Phi 2})^2]}{2\lambda}~,  \\
m_{S_2}^2 &\simeq& v_R^2\frac{(\lambda_{\eta\Phi 2}-\lambda_{\eta\Phi 1})}{2}  ~,\nn\\
m_{S_R}^2 &\simeq& v^2_R(2 \lambda) \nn~,
\end{eqnarray}
which are associated with the physical fields whose main contribution come from the fields $S_1, S_2$ and $S_R$, respectively. The lighter among the three real scalar fields is therefore identified with the SM Higgs field, with mass $m_{S_1} \simeq 125$~GeV, whereas the other two are heavier Higgses.

\section{The gauge boson mass spectrum}\label{sec.4}

The gauge bosons become massive via the Higgs mechanism once the scalar fields acquire non-vanishing vevs. Their masses can be obtained from the following terms
\be 
\mathcal{L} = (D^\mu \eta_L)^\dagger(D_\mu \eta_L) + (D^\mu \eta_R)^\dagger(D_\mu \eta_R) + (D^\mu \Phi)^\dagger(D_\mu \Phi) + (D^\mu \sigma)^*(D_\mu \sigma)~.
\ee 
Because $\sigma$ is a gauge singlet, $D_\mu \sigma \to \partial_\mu \sigma$, it does not contribute to the gauge field masses. 

When the scalar fields acquire vevs, as described in eq. (\ref{scdec}) with $v_L=0$, the charged gauge bosons $\tilde{W}_{L(R)}^\pm = (W_{L(R)}^1 \mp i W_{L(R)}^2)/\sqrt{2} $ mix with each other. The mass eigenstates, $W_{L}^\pm$ and $W_{R}^\pm$, can be defined as the linear combinations below
\be \label{cgb}
\begin{pmatrix}
W_{L\mu}^+ \\ W_{R\mu}^+  
\end{pmatrix}
=
\begin{pmatrix}
\cos\zeta & \sin\zeta \\ - \sin\zeta & \cos \zeta
\end{pmatrix}
\begin{pmatrix}
\tilde{W}_{L\mu}^+ \\ \tilde{W}_{R\mu}^+  
\end{pmatrix}~, ~~\mbox{with}~~ \tan(2\zeta) = -\frac{4v_{1}v_{2}}{v_R^2}~,
\ee 
and the associated masses are
\be \label{mwlr}
m_{W_{R(L)}}^2 = \frac{1}{4} g^2 \left[ v^2 +\frac{ v_R^2}{2}\pm \frac{1}{2}\sqrt{16 ( v_{1} v_{2})^2+ v_R^4}\right]~.
\ee 
In the limit $v_R\gg v_1 \gg v_2$, with $v^2=v_1^2+v_2^2\simeq v_1^2$, the masses in eq. (\ref{mwlr}) become
\bea \label{Wmass}
m_{W_L}^2 \simeq \frac{g^2}{4}v^2 ~~~\mbox{and}~~~m_{W_R}^2 \simeq \frac{g^2}{4} (v_R^2 + v^2)~,
\eea
From the expressions above, we conclude that $v = v_{EW} \approx 246$ GeV and $W_L$ is identified with the SM charged vector boson. Furthermore, taking, for instance, $v_R = 15$~TeV, $v_1\sim v_{EW} = 246$ GeV and $v_2 \sim 1$ GeV, we have that $m_{W_R}\simeq 4.9$ TeV, and the mixing angle is very small: $|\zeta| \simeq 2.2 \times 10^{-6}$. For these values, $m_{W_R}$ is above the current lower bounds coming from new particle searches at the LHC \cite{Sirunyan:2018pom,Aaboud:2018spl}.

Let us consider now the real gauge bosons $W_{L\mu}^3, W_{R\mu}^3$ and $B_\mu$. They mix among themselves and give rise to one massless and two massive neutral gauge bosons. The massless field is the photon field $A_\mu$; one of the massive fields is the SM Z boson, while the other is a heavy neutral gauge boson $Z^\prime$:
\begin{eqnarray}
\begin{pmatrix}
A_\mu \\ Z_\mu \\ Z^\prime_\mu
\end{pmatrix}
= \begin{pmatrix}
\sin(\theta_W) & \sin(\theta_W) & \sqrt{\cos(2 \theta_W)}\\ \cos(\theta_W) & -\sin(\theta_W) \tan(\theta_W) & - \tan(\theta_W) \sqrt{\cos(2\theta_W)}  \\ 0 & -\sec(\theta_W)\sqrt{\cos(2 \theta_W)} & \tan(\theta_W) 
\end{pmatrix}
\begin{pmatrix}
W^3_{L \mu} \\ W^3_{R \mu} \\ B_\mu
\end{pmatrix}~,
\end{eqnarray}
where $\theta_W$ is the Weinberg angle, for which $\sin^2\theta_W\simeq 0.231$. The masses are given by
\begin{eqnarray}\label{Zmass}
m_{Z^\prime}^2 &=& \frac{g^2 v_R^2}{4(1-\tan^2\theta_W)} + m_Z^2 \cos(2\theta_W)~,\\
m_Z^2 &=& \frac{g^2 v^2}{4 \cos^2\theta_W}~,  \nn
\end{eqnarray}
and the mixing angle between the massive fields is defined as 
\begin{equation}
\tan(2 \varphi) = \frac{2 m_Z^2 \sqrt{\cos(2 \theta_W)}}{m_{Z^\prime}^2-m_Z^2}~.
\end{equation}
From eqs. (\ref{Wmass}) and (\ref{Zmass}), we have the relation $m_{Z^\prime}\simeq 1.2\times m_{W_R}$. When assuming, for example, $v_R = 15$ TeV, we find that $m_{Z^\prime}\simeq 5.8$ TeV and $\varphi \simeq 1.79\times 10^{-4}$.

\section{Fermion masses and mixing}\label{sec.5}

Considering the field content and symmetries presented in section \ref{sec.2}, the following renormalisable Yukawa terms can be written down
\begin{equation}
\label{ReYuk}
\begin{split}
-\mathcal{L}_Y  & =\, y_{\alpha \beta}^{q} \, \overline{Q_{\alpha L}} \, \Phi \, Q_{\beta R} + y_{\alpha \beta}^{l}\, \overline{L_{\alpha L}}\,\Phi \, L_{\beta R} \\
  & + y_{\alpha \beta}^* \,\overline{L_{\alpha L}} \,\eta_L\, S_{\beta R} + y_{\alpha \beta}\, \overline{L_{\alpha R}}\, \eta_R\, (S_{\beta R})^c + h.c.~,
\end{split}
\end{equation}
with $y_{\alpha \beta}^{l,q} = (y_{\alpha \beta}^{l,q})^T$. The first term provides all quarks with masses when the neutral fields in $\Phi$ acquire nonvanishing vevs. However, in this case both up-type and down-type quark mass matrices are proportional to $y^q_{\alpha\beta}$ and, as a consequence, these matrices can be diagonalised simultaneously. Thus, the corresponding tree-level CKM matrix does not mix quark flavours, disagreeing with the experimental picture. This is a direct consequence of having an extra symmetry that distinguishes $\Phi$ from its charged conjugated $\tilde{\Phi}$ and, as such, forbids the operator $\overline{Q_{L}} \, \tilde{\Phi} \, Q_{R}$ from appearing in the tree-level Lagrangian above. In our case, this symmetry is $U(1)_{PQ}$. Additionally, since the eq. (\ref{ReYuk}) does not present a Majorana mass term for the neutral fermion singlets $S_{\alpha R}$, some neutrinos remain massless. 

These two issues can be dealt with by considering the following nonrenormalisable operators involving the scalar singlet $\sigma$
\begin{eqnarray}\label{HOYuk}
-\mathcal{L}_Y^{\sigma} &=& h_{ \alpha \beta}^{q} \, \left( \frac{\sigma}{\Lambda}\right)\,\overline{Q_{\alpha L}}\, \tilde{\Phi}\, Q_{\beta R} + h_{ \alpha \beta}^{l} \, \left( \frac{\sigma}{\Lambda}\right)\,\overline{L_{\alpha L}}\, \tilde{\Phi}\, L_{\beta R} \\
&&+ \frac{1}{2}h_{\alpha \beta}\, \left( \frac{\sigma}{\Lambda} \right)^n\, \sigma \, \overline{S_{\alpha R}}\, (S_{\beta R})^c+ h.c.~,\nn
\end{eqnarray}
with $y_{\alpha \beta}^{l,q} = (y_{\alpha \beta}^{l,q})^T$, $h_{\alpha \beta} =h^T_{\alpha \beta}$, $n$ is an integer, kept arbitrary by now, associated with the PQ symmetry, see Table \ref{t1}, and $\Lambda$ is a large mass scale suppressing the higher-dimensional operators. It is worth pointing out that since neither $S_{\alpha R}$ nor $\sigma$ carry lepton number, Lepton number violation occurs when $\eta_R$ gets a vev.

\subsection{Quark masses: suppressed quark mixing}

With the introduction of the higher-dimensional operators in eq. (\ref{HOYuk}) in addition to those in eq. (\ref{ReYuk}), the up and down-type quark mass matrices, written according to $\overline{{\bf u}}_L \,{\bf M}_u \,{\bf u}_R$ and $\overline{{\bf d}}_L \,{\bf M}_d \,{\bf d}_R$ with ${\bf u}_{L(R)} = 
(u, c, t)_{L(R)}^T$ and ${\bf d}_{L(R)} = 
(d, s, b)_{L(R)}^T$,  become
\bea\label{qmass} 
{\bf M}_{u} &=&  {\bf y}^{q}\, \frac{v_1}{\sqrt{2}} + {\bf h}^{q} \, \frac{\epsilon v_2}{\sqrt{2}}~,\\
{\bf M}_{d} &=&  {\bf y}^{q} \,\frac{v_2}{\sqrt{2}} + {\bf h}^{q}  \,\frac{\epsilon v_1}{\sqrt{2}}~,\nn
\eea
with $\epsilon \equiv v_\sigma/(\sqrt{2}\Lambda)\ll 1$.
Such mass matrices are no longer proportional to each other due to the contributions suppressed by $\epsilon$. In this way our model predicts that the non-trivial mixing angles of the new CKM matrix vanish when the suppression factor $\epsilon$ approaches zero, providing thus a natural explanation to the ``flavour alignment" puzzle.

The effective operators in eq. (\ref{HOYuk}), leading to a small departure from alignment, are expected to come from a UV-complete theory upon the integration of heavy degrees of freedom. In ref. \cite{Dev:2018pjn}, for example, a UV-complete model featuring, among other new fields, heavy coloured scalars explains the observed small departure from flavour alignment as a result of it being generated at loop level only.

In general, the matrices in eq. (\ref{qmass}) can be diagonalised following the transformations ${\bf u}_{L(R)} \to {\bf U}_{L(R)} {\bf u^\prime}_{L(R)}$, ${\bf d}_{L(R)} \to {\bf D}_{L(R)} {\bf d^\prime}_{L(R)}$, where ${\bf U}_{L(R)}$ and ${\bf  D}_{L(R)}$ are unitary matrices, and the primed fields represent the physical states. The imposition of $\mathcal{C}$ as the LR symmetry, as described by eq. (\ref{cconjf}), implies that the left and right transformations are not independent but related via ${\bf U}_R = {\bf U}^*_L {\bf K}_u^*$ and ${\bf D}_R = {\bf D}^*_L {\bf K}_d^*$, with ${\bf K}_u$, ${\bf K}_d$ diagonal matrices of phases. Also, the CKM matrix can be defined as ${\bf V}_{CKM} \equiv {\bf V}_{L}={\bf U}^\dagger_L {\bf D}_L$. 

There is enough freedom in the parameter space to easily accommodate all the quark masses and mixing. For instance, the values below can be used as benchmarks to fit the data: 
\bea \label{v1v2e}
v_1 \simeq 245.927~\mbox{GeV}, ~~v_2 \simeq 5.991~\mbox{GeV},~~\epsilon\simeq 3.39\times10^{-4}~,
\eea
and the matrices ${\bf D}_L$, ${\bf y}^q$ and ${\bf h}^q$ are given in the Appendix \ref{AppB}.

\subsection{Lepton masses: inverse seesaw mechanism}\label{neum}

For the charged leptons, we find
\bea\label{fermass} 
{\bf M}_{l} &=&  {\bf y}^{l} \,\frac{v_2}{\sqrt{2}} + {\bf h}^{l} \, \frac{\epsilon v_1}{\sqrt{2}}~.
\eea
We can take this matrix to be diagonal without loss of generality. Moreover, from eq. (\ref{v1v2e}) we see that $\epsilon v_1/v_2 \ll 1$, and assuming, for simplicity, $ h^l_{ij}\simeq y^l_{ij}$, we can take
\begin{equation}\label{yl}
{\bf y}^l \simeq \frac{\sqrt{2}}{v_2} \mbox{diag}(m_e,~m_\mu,~ m_\tau)~.
\end{equation}

Let us now investigate how neutrinos become massive in our model. The nine neutral fermions can be grouped into the following flavor basis ${\bf N}_{L}=(\nu_{\alpha L}, (N_{\alpha R})^c, (S_{\alpha R})^c)$. When the scalar fields get their respective vevs, the interaction terms in eqs. (\ref{ReYuk}) and (\ref{HOYuk}) give rise to the following $9\times 9$ neutrino mass matrix, using the convention $(1/2) \overline{({\bf N}_L)^c}\, \mathcal{M}_\nu\, {\bf N}_L$,
\be\label{issmatrix}
\mathcal{M}_\nu = \begin{pmatrix}
 0 & {\bf m_{D}} & 0 \\
 {\bf m_D}& 0 &{\bf m_R} \\
 0 & ({\bf m_R})^T & \textbf{ {\boldmath $\mu$} }
\end{pmatrix}~,
\ee 
with $\sqrt{2}\,{\bf m_{D}} = ({\bf y}^{l})^* \,v_{1}+ ({\bf h}^{l})^* \,\epsilon\, v_{2}   $, $\sqrt{2}\, {\bf m_R} =  {\bf y}\,v_R  $, $\sqrt{2}\,\textbf{{\boldmath $\mu$}} = {\bf h} \,\epsilon^n\, v_\sigma $. Due to our choice of $\mathcal{C}$ as the LR symmetry, we have that $({\bf m_D})^T = {\bf m_D}$.

When considering the solution in eq. (\ref{yl}), ${\bf m_D}$ can also be taken diagonal and given in terms of the charged lepton masses
\begin{equation}\label{mD}
{\bf m_D} = \frac{v_1}{v_2}\,\mbox{diag}(
m_e,~m_\mu,~m_\tau)~.
\end{equation}
Moreover, there is enough freedom to transform $S_{\alpha R}$ to make $\textbf{{\boldmath $\mu$}}$ diagonal
\be \label{mu}
\textbf{ {\boldmath $\mu$}}= \frac{\epsilon^n v_\sigma}{\sqrt{2}}\,{\bf h}= \frac{\epsilon^n v_\sigma}{\sqrt{2}}\,\mbox{diag}(h_1,~h_2,~h_3)~.
\ee
For $n$ large, the following hierarchy is obtained ${\bf m_R} \gg {\bf m_D} \gg \textbf{{\boldmath$\mu$}}$, which together with the texture of the mass matrix in eq. (\ref{issmatrix}) allows for the inverse seesaw to take place.

Neutrino masses are then obtained by diagonalising the mass matrix in eq. (\ref{issmatrix}) with the help of the unitary matrix $\mathcal{U}$ which can divided into two matrices: $\mathcal{U}=\mathcal{U}_1 \mathcal{U}_2$. Making use of the fact that ${\bf m_R} \gg {\bf m_D} \gg \textbf{{\boldmath$\mu$}}$, we can write $\mathcal{U}_1$ as 
\be \label{U1}
\mathcal{U}_1 \simeq \begin{pmatrix} {\bf 1} & -{\bf ab} & {\bf a} \\
{\bf b}^\dagger{\bf a}^\dagger & {\bf 1} & {\bf 0} \\
-{\bf a}^\dagger & {\bf 0} & {\bf 1}
\end{pmatrix}~,
\ee 
where ${\bf a} = {\bf m_D} ({\bf m_R}^\dagger)^{-1}$ and ${\bf b} = \textbf{{\boldmath$\mu$}} ({\bf m_R}^*)^{-1}$, and all the elements of the matrices above represent $3\times 3$ matrices.
When $\mathcal{U}_1$ acts on $\mathcal{M}_\nu$, the mass matrix is block-diagonalised 
\be 
\mathcal{M}^1_\nu = \mathcal{U}_1^T\,\mathcal{M}_\nu\, \mathcal{U}_1 = \begin{pmatrix}
{\bf M_\nu} & {\bf 0} &  {\bf 0}\\
 {\bf 0} & {\bf M_{11}^h} & {\bf M_{12}^h} \\
 {\bf 0} & {\bf M_{12}^h} & {\bf M_{22}^h }
\end{pmatrix}~,
\ee 
where ${\bf M}_\nu$ is the mass matrix of the light neutrinos $\nu_{iL}$ with masses $m_i$, while the $3 \times 3$ entries ${\bf M^h_{ij}}$ form the $6\times 6$ mass matrix of the heavy neutrinos $N_{iR}$ and $S_{iR}$ with masses $M_{N_i}$ and $M_{S_i}$. 
At leading order, the lightest states, i.e. the active neutrinos, will have the following mass matrix
\be\label{mnuISS}
{\bf M}_\nu \simeq {\bf m_D} ({\bf m_R}^T)^{-1} \textbf{{\boldmath$\mu$}}({\bf m_R})^{-1} {\bf m_D}~.
\ee
Finally, for the last step, we define $\mathcal{U}_2$ as
\be \label{U2}
\mathcal{U}_2 = \begin{pmatrix}
{\bf U} & {\bf 0} & {\bf 0}\\
{\bf 0} & {\bf V_{11}} & {\bf V_{12} }\\
{\bf 0} & {\bf V_{21}} & {\bf V_{22}} 
\end{pmatrix}~,
\ee
where ${\bf U}$ is identified with the unitary PMNS mixing matrix that diagonalises ${\bf  M}_\nu$, while the $6 \times 6$ matrix formed by the ${\bf V_{ij}}$ matrices diagonalises the heavy neutrino mass matrix so that\footnote{In this approximation, we have neglected higher-order contributions in $\mathcal{U}$ which would eventually introduce nonunitarity corrections to the matrix that diagonalises eq. (\ref{mnuISS}). At leading order, such corrections are of order ${\bf a} {\bf a}^\dagger$.}
\be 
\mathcal{M}^{(diag)}_\nu = \mathcal{U}^T\,\mathcal{M}_\nu\, \mathcal{U}~.
\ee 

Considering $v_R =15$ TeV, $v_\sigma = 10^{11}$ GeV, and the benchmark values in eq. (\ref{v1v2e}), we have
\begin{equation}\label{mnun}
{\bf M}_\nu \simeq  \left[{\bf d_1} ({\bf y}^T)^{-1} {\bf h}({\bf y})^{-1}{\bf d_1}\right]\times(0.33437 \times 3.3925^n)\times 10^{4(4-n)}~\mbox{eV}~,
\end{equation}
with ${\bf d_1}  = \mbox{diag}(2.8\times 10^{-4},~5.946\times10^{-2},~1)$. Sub-eV neutrino masses follow naturally from eq. (\ref{mnun}) when $n=4$ without the need for any unnaturally small coupling in ${\bf y}$ or ${\bf h}$. On the other hand, for $n<4$, sub-eV masses require ${\bf h}$ to be very small when compared to ${\bf y}$; whereas for $n>4$, ${\bf y}$ needs to be more and more suppressed with respect to ${\bf h}$. 

We set $n=4$ and, in the Appendix \ref{AppC}, provide a point in the parameter that shows that eq. (\ref{mnun}) can be easily used to fit the current neutrino data (masses and mixing). The solution presents normal mass ordering of the active neutrinos which is currently favoured over the inverted ordering case~\cite{Vagnozzi:2017ovm, Esteban:2018azc, Dolinski:2019nrj}. 

Finally, we would like to stress out the importance of the PQ symmetry for the generation of neutrino masses. The smallness of the active neutrino masses follow from their dependence on the PQ scale, $v_\sigma$, through the suppression factor $\epsilon$, rather than from a strong dependence on the largeness of the scale $v_R$, as it is common when the type-I seesaw mechanism is in place. Therefore, all the current neutrino data can be successfully reproduced by our model even if the scale $v_R$ is only high enough to evade the constraints from new vector boson searches. This is in contrast to ref. \cite{Dev:2018pjn}, where the type-I seesaw mechanism is realised and the authors argue that only for values of $v_R\geq 50$ TeV can the current neutrino data be fitted.

\section{Neutrinoless double beta decay} \label{sec.6}

New contributions to $0\nu2\beta$ arise in the LR model in processes where $W_R$ is exchanged instead of $W_L$ \cite{Hirsch:1996qw, Maiezza:2010ic, Barry:2013xxa, Dev:2014xea}. If the LR symmetry is broken at a very high scale, the new contributions can be neglected in favour of the standard one which is proportional to the Majorana effective mass $\langle m_{ee} \rangle$. In the present case, however, with $v_R = 15$ TeV, $W_R$ as well as the heavy neutrinos will get masses around the TeV scale, so that the contributions to $0\nu2\beta$ involving such particles may become relevant and therefore need to be checked.

In order to estimate the non-standard contributions, let us first write down, in the basis where the charged leptons are diagonal, the interactions in the lepton sector mediated by the charged gauge bosons $W_L$ and $W_R$, 
\bea\label{CC}
\mathcal{L}_{CC} &=& \frac{g}{\sqrt{2}} \left( \overline{l}_{\alpha L}\, \gamma^\mu \, \nu_{\alpha L} \,\tilde{W}_{L \mu}^-+ \overline{l}_{\alpha R} \, \gamma^\mu\, N_{\alpha R}\, \tilde{W}_{R \mu}^- + h.c.\right) \\
&\simeq& \frac{g}{\sqrt{2}} \left\{ \overline{l}_{\alpha L}\, \gamma^\mu \left[ {\bf U}_{\alpha i} \nu_{i L}+({\bf a}{\bf V_{21}})_{\alpha i} (N_{i R})^c +({\bf a}{\bf V_{22}})_{\alpha i} (S_{i R})^c\right]( W_{L \mu}^- - \zeta W_{R \mu}^-) \right.\nn \\
 & +& \left.  \overline{l}_{\alpha R} \, \gamma^\mu\, \left[({\bf b}^\dagger{\bf a}^\dagger {\bf U})^*_{\alpha i} (\nu_{i L})^c+({\bf V_{11}})^*_{\alpha i} N_{i R} +({\bf V_{12}})^*_{\alpha i} S_{i R} \right]\, (W_{R\mu}^- + \zeta W_{L\mu}^-) + h.c.\right\},\nn
\eea 
where from the first to the second line, we have used eqs. (\ref{cgb}), (\ref{U1}) and (\ref{U2}) to write the flavour neutrinos and gauge bosons in terms of their corresponding mass states. For the processes in which we are interested, only interactions with (anti-)electrons are relevant, thus we choose $\alpha=e$. 
\begin{figure}[t!]
\centering
\includegraphics[scale=.265]{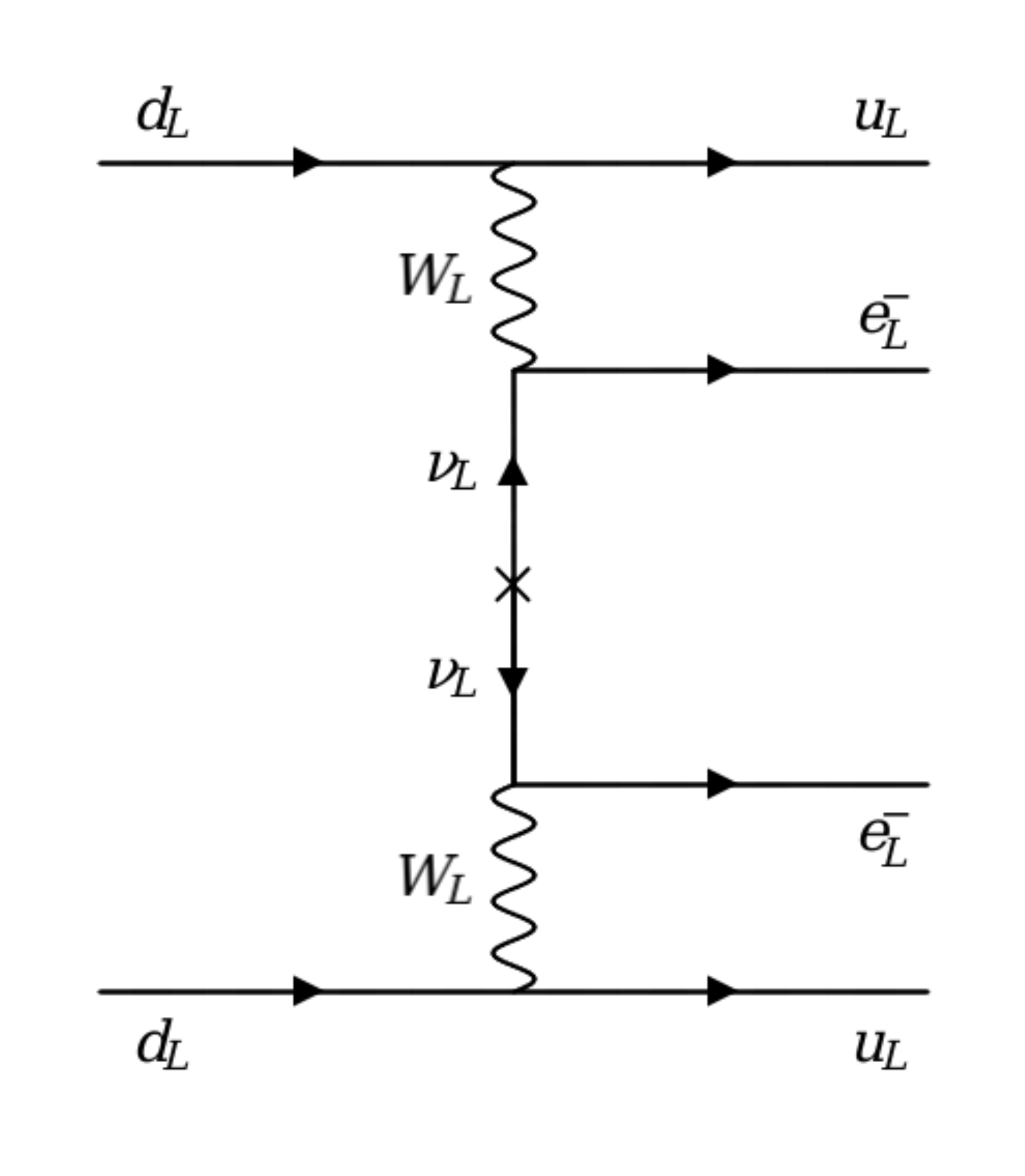}
\includegraphics[scale=.265]{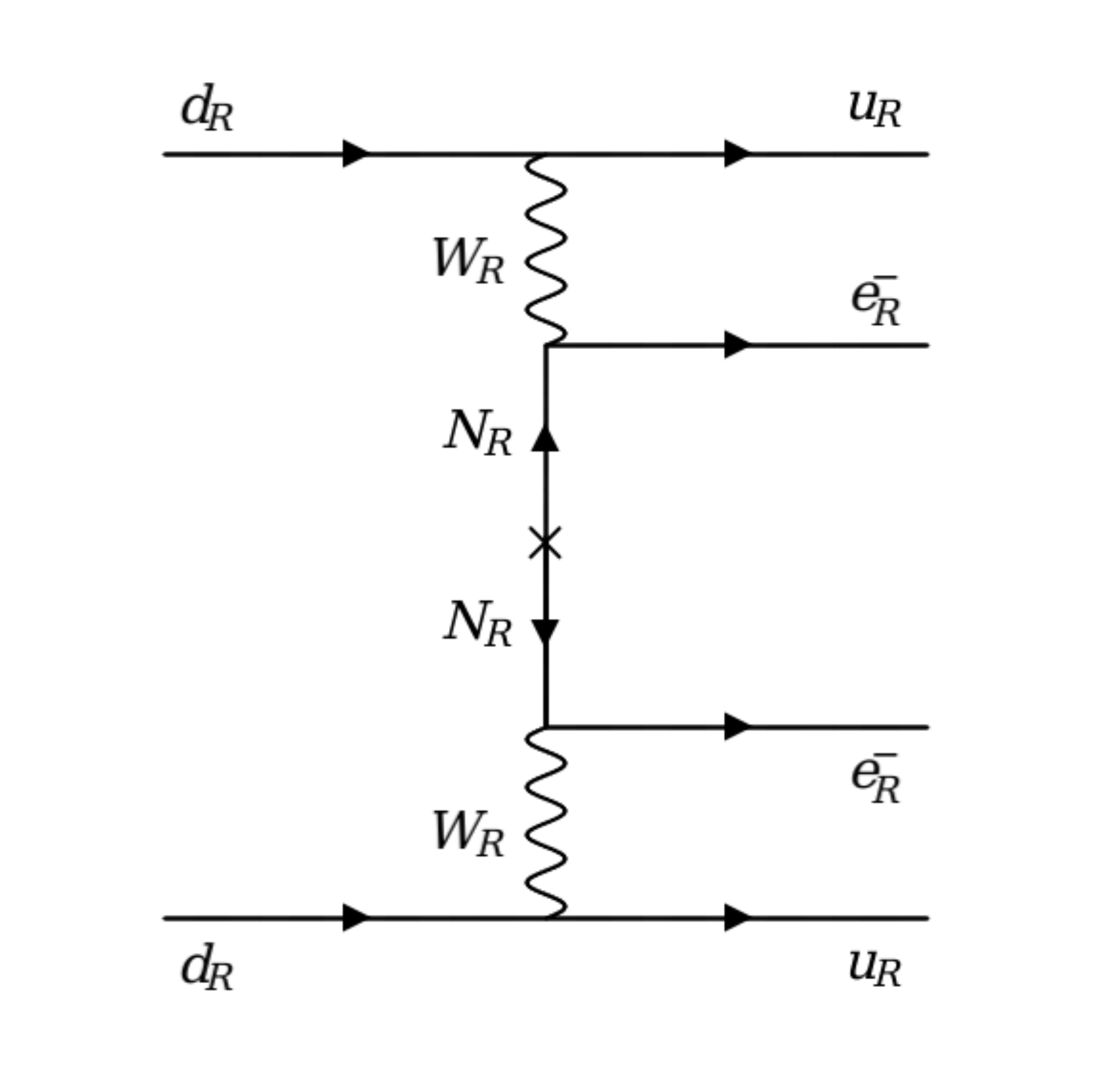}
\includegraphics[scale=.265]{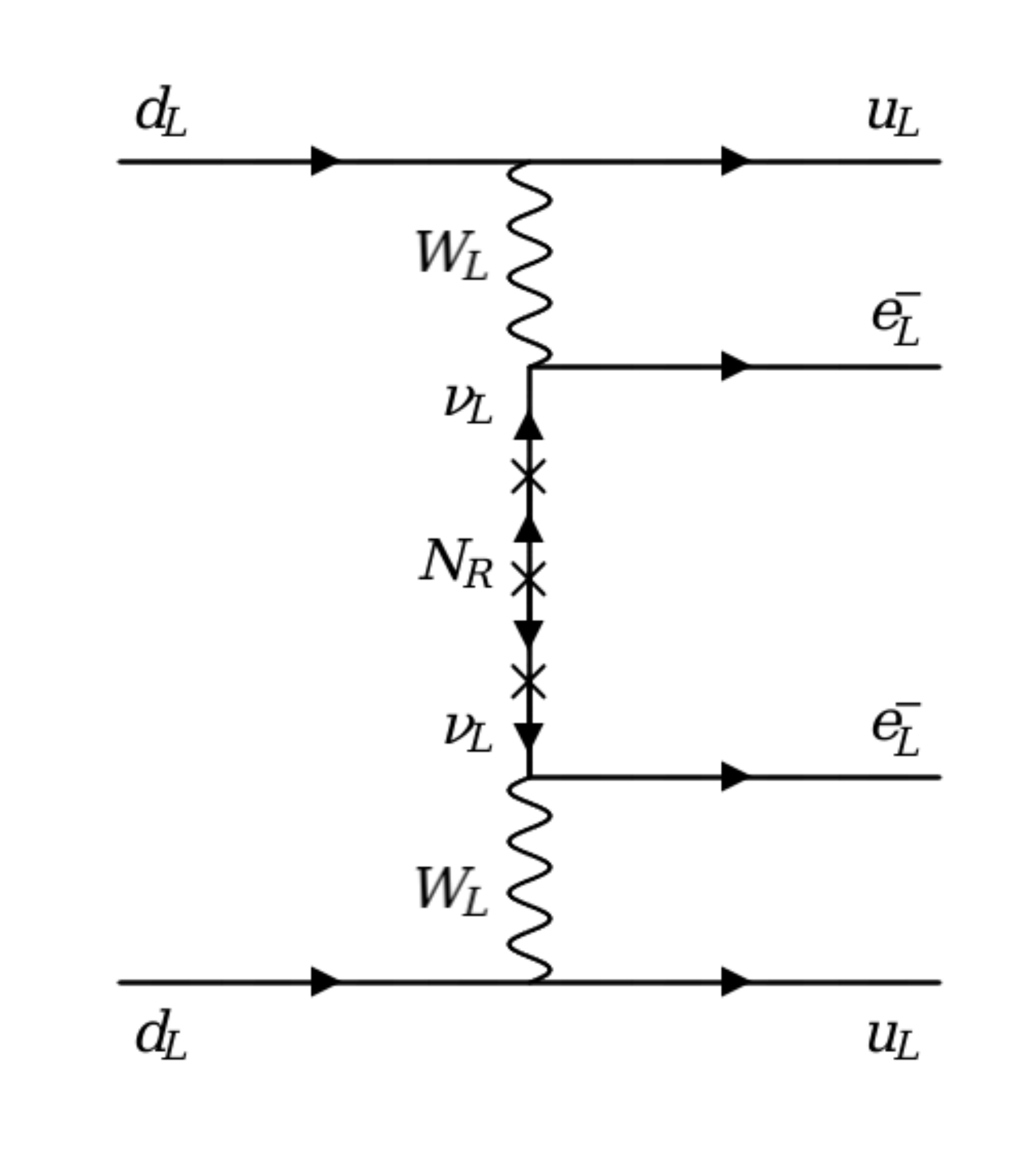}
\caption{$0\nu2\beta$ diagrams leading to the standard contribution, its right counterpart, and the light-heavy neutrino mixing contribution, respectively.}
\label{fig1}
\end{figure}

\begin{figure}[t!]
\centering
\includegraphics[scale=.265]{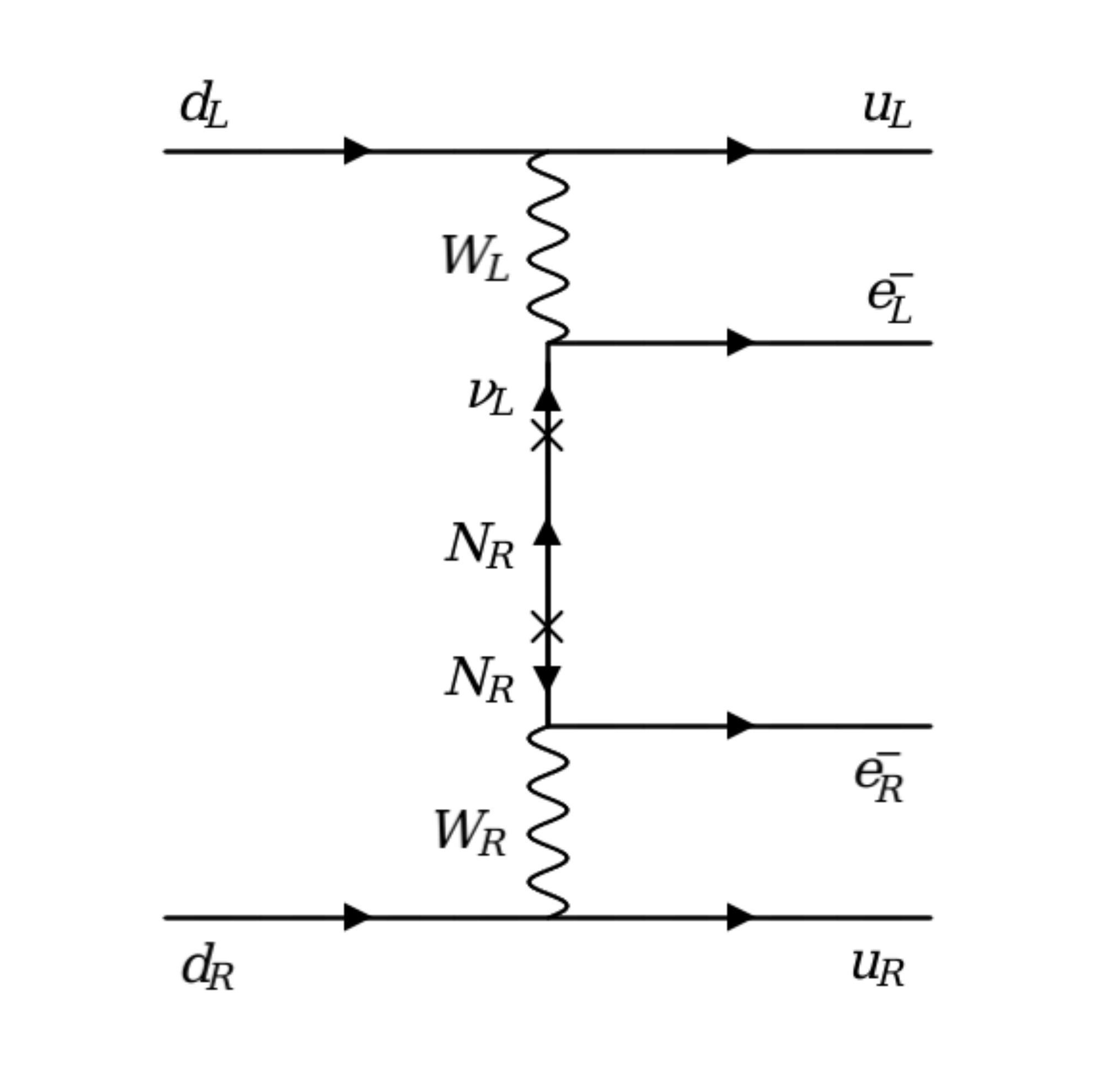}\hspace{2cm}
\includegraphics[scale=.265]{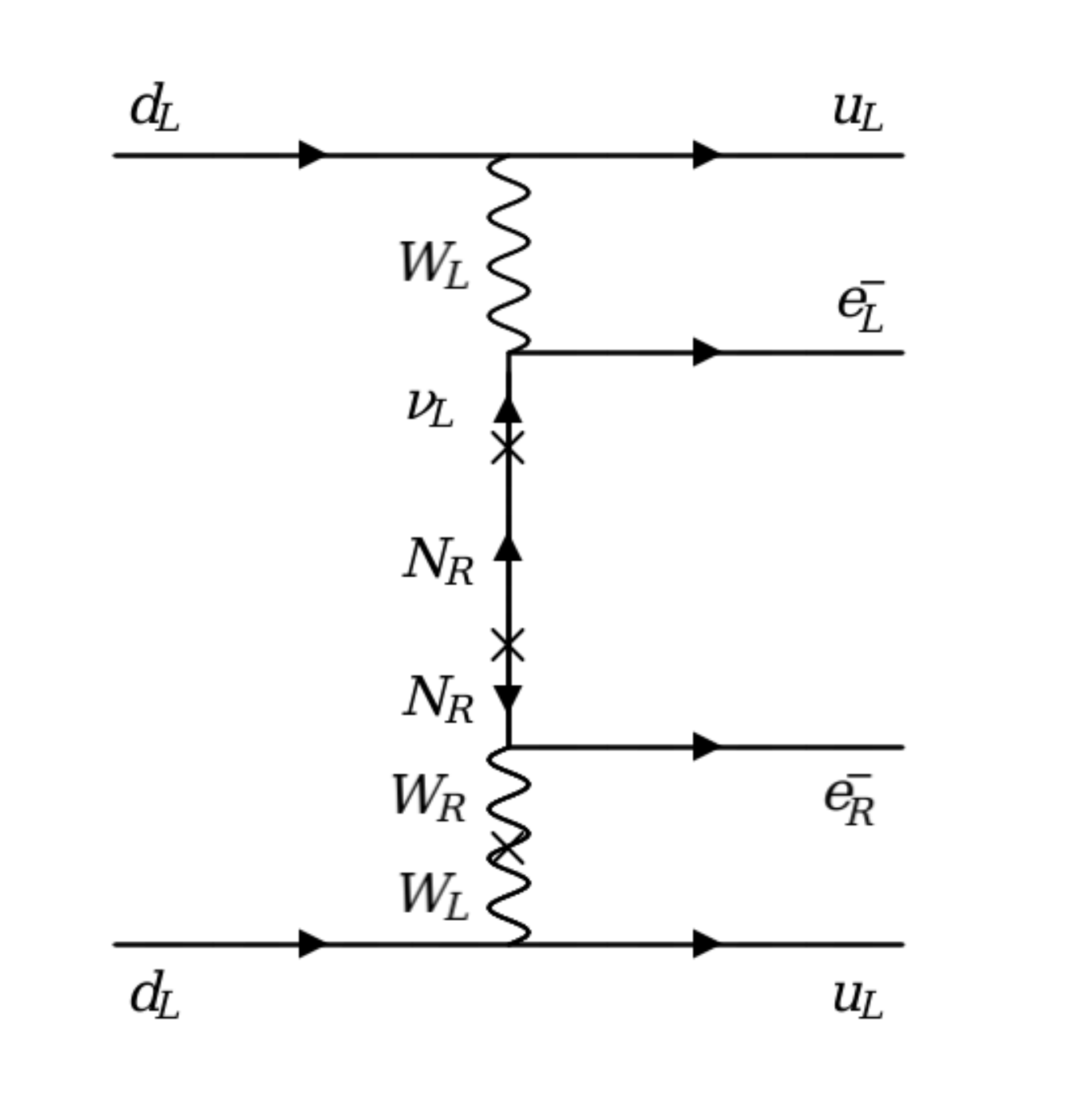}
\caption{Other possible contributions to  $0\nu\beta\beta$ for which the final electrons present opposite chiralities. }
\label{fig2}
\end{figure}

The standard contribution to $0\nu2\beta$ follows from the exchange of $W_L$ gauge bosons and involve the Majorana masses of the left-handed neutrinos, as shown in the first diagram in Fig. \ref{fig1}. The LR symmetry implies that the diagram in the centre of Fig. \ref{fig1} also needs to be taken into account as the right counterpart of the standard contribution. The mixing between light and heavy neutrinos introduces another possibility shown in the last diagram in Fig. \ref{fig1}. Contributions involving final electrons with opposite chiralities are also possible, as we can see in Fig. \ref{fig2}.

The leading contributions coming from the diagrams in Figs. \ref{fig1} and \ref{fig2} can be expressed in terms of the amplitudes below, respectively,
\bea \label{amplitudes}
\mathcal{A}_\nu &\simeq& G_F^2\,\frac{\langle m_{ee}\rangle}{q^2}~,\\
\mathcal{A}^R_{N_R} &\simeq& G_F^2\, \left(\frac{m_{W_L}}{m_{W_R}}\right)^4  \sum_{i}\left[\frac{({\bf V_{11}^*})_{ei}^2}{M_{N_i}}+\frac{({\bf V_{12}^*})_{ei}^2}{M_{S_i}}\right] ~,\nn\\
\mathcal{A}^L_{N_R} &\simeq& G_F^2\, \sum_i\left[\frac{({\bf a} {\bf V_{21}})_{ei}^2}{M_{N_i}}+\frac{({\bf a} {\bf V_{22}})_{ei}^2}{M_{S_i}}\right]   ~,\nn\\
\mathcal{A}_\lambda &\simeq& G_F^2\, \left(\frac{m_{W_L}}{m_{W_R}}\right)^2\,\sum_{i}\frac{{\bf U}_{ei} {({\bf b}^\dagger {\bf a}^\dagger {\bf U})}_{ei}^*}{q}\,,\nn\\
\mathcal{A}_\eta &\simeq& G_F^2\, \tan(\zeta)\,\sum_{i}\frac{{\bf U}_{ei} {({\bf b}^\dagger {\bf a}^\dagger {\bf U})}_{ei}^*}{q}\,,\nn
\eea 
where $G_F$ is the Fermi constant, $\langle m_{ee}\rangle = \sum_i{\bf U}_{ei}^2 m_i$ the effective Majorana mass, $q \simeq 100$ MeV the typical momentum exchange of the process, and $M_{N_i}$ and $M_{S_i}$ the masses of the heavy neutrinos $N_{iR}$ and $S_{iR}$, respectively.

For the benchmark point in the parameter space given in the Appendix \ref{AppC}, the active and heavy quasi-Dirac neutrino pair masses, $m_i$ and $m^h_i$, respectively, are:
\bea\label{neutmasses}
m_i &=& (0.90,\, 1.24,\, 5.139)\times 10^{-2}~\mbox{eV}~\\
m^h_i &=& (0.532,\, 2.750,\,14.78)~\mbox{TeV}~,\nn
\eea 
and the effective Majorana mass is
\be\label{effmm}
\langle m_{ee} \rangle = 2.14\times10^{-3}~\mbox{eV}~,
\ee
which is a typical value for $\langle m_{ee} \rangle$ in the case of normal mass ordering.\footnote{ Several other solutions with similar or smaller $\langle m_{ee} \rangle$ values can also be easily found.} The mixing between the charged gauge bosons $W_L$ and $W_R$ is $\zeta\simeq -1.3\times 10^{-5}$.
By calculating and comparing all the contributions we have found that the standard $\mathcal{A}_\nu$ is by far the dominant one. Whereas among the non-standard contributions, $\mathcal{A}_\lambda$ is the largest but still very small when compared to $\mathcal{A}_\nu$:  $|\mathcal{A}_\lambda/\mathcal{A}_\nu| \simeq 1.26\times 10^{-3} $. Therefore, at least for this specific solution, since the dominant contribution is the standard one governed by small effective Majorana mass in eq. (\ref{effmm}), no evidence for $0\nu2\beta$ is expected to be observed in the next generation experiments which will scan the region $|\langle m_{ee} \rangle|\sim (0.01 - 0.05)$ eV \cite{Tanabashi:2018oca,Dolinski:2019nrj}.

\section{Lepton flavour violation and heavy neutrino constraints}\label{sec.7}

We now discuss two important phenomenological implications of our model: lepton flavour violation and heavy neutrino constraints. We start by considering lepton flavour violation through the decays of a charged lepton into a different charged lepton plus a photon: $l_i \to l_j \gamma$, since SM extensions featuring TeV scale neutrino mass mechanisms, like ours, can induce large branching ratios for such decays. Among the possible decays, the constraint on the muon decay is the strongest one with $BR(\mu^+\to e^+ \gamma) < 4.2\times 10^{-13}$ \cite{TheMEG:2016wtm}. Whereas the current upper limits for the branching ratio associated with tau decays, {\it i.e.} $\tau\to \mu \gamma$ and $\tau \to e \gamma$, are of order of $10^{-8}$~\cite{Aubert:2009ag}. The branching ratio for these processes can be written as \cite{Barry:2013xxa}
\be\label{brex}
BR(l_i \to l_j \gamma) = \frac{3 \alpha_{em}}{2 \pi}\left(|\left(G_L^\gamma\right)_{ij}|^2+|\left(G_R^\gamma\right)_{ij}|^2\right)~,
\ee
where $G_L^\gamma$ and $G_R^\gamma$ are explicitly given in the Appendix \ref{AppD}. Using our solution in Appendix \ref{AppC}, we calculate these branching ratios for the three processes to find
\bea\label{BR}
BR(\mu\to e \gamma ) &\simeq& 2.1\times 10^{-13}~,\\
BR(\tau \to \mu \gamma ) &\simeq& 6.5\times 10^{-13}~,\nn\\
BR(\tau \to e \gamma ) &\simeq& 9.8\times 10^{-14}~.\nn
\eea
The dominant contributions come from the terms in $G^\gamma_L$, as defined by Eq. (\ref{GLR}), except for those proportional to $|\zeta|^2$ which are naturally too suppressed. It is clear from Eq. (\ref{BR}) that our predictions for both tau decays are still quite far from the current experimental limits. For the muon decay, on the other hand, the value we found is just below the current upper limit. With the forthcoming MEG II experiment that aims at improving the current $BR(\mu \to e \gamma)$ limit by an order of magnitude, our prediction could be soon put to the test. 

In addition to lepton flavour violation, the presence of heavy neutrinos that mix with the active ones can lead to other distinctive experimental signatures. For example, collider searches for events with three charged leptons (plus an active neutrino that escapes detection) arising from the decay of heavy neutrinos can be used to constrain how the latter can mix with the active neutrinos. Searches for this signature have been performed using the CMS detector at the LHC considering heavy neutrino masses from $1$ GeV to $1.2$ TeV and decays mediated by $W_L$ \cite{Sirunyan:2018mtv}. According to our solution in Eq.~(\ref{neutmasses}), only the lightest two among the heavy Majorana neutrinos, forming a quasi-Dirac pair with mass just above $530$ GeV, are within the analysed mass range. The results presented in Ref.~\cite{Sirunyan:2018mtv} require the mixing between these heavy states and the first two active neutrinos to be smaller than around $10^{-1}$. The relevant mixing matrices for our benchmark solution are shown in Eq.~(\ref{lhmix}), from which we can conclude that the above constraint is easily satisfied since the largest entries in Eq.~(\ref{lhmix}) are no larger than $10^{-3}$.

Finally, it is worth mentioning that if we had considered different benchmarks, other processes could become more relevant to assess the viability of our results. For instance, if at least one of the heavy neutrino masses was close to $125$ GeV, the decay of the Higgs boson into such a state plus an active neutrino would be kinematically allowed, which would lead to constraints on the corresponding Dirac Yukawa couplings. For discussions on this topic, see, for example, Refs.~\cite{BhupalDev:2012zg, Cely:2012bz, Das:2017zjc}.

\section{Conclusions}\label{sec.8}

In this work, we have presented a minimal left-right model with a Peccei-Quinn symmetry and shown how it can help us to understand the origin of two intriguing features, namely, the approximately diagonal structure of the quark mixing (CKM) matrix and the smallness of neutrino masses. 

The scalar sector is composed of a gauge singlet, a $SU(2)_R$ and a $SU(2)_L$ doublet, and a bi-doublet: $\sigma$,  $\eta_R$,  $\eta_L$ and $\Phi$, respectively. Symmetry breaking occurs in three stages. Firstly, when the singlet acquires a vev, $\langle \sigma \rangle \propto v_\sigma = 10^{11}$ GeV, the PQ symmetry is spontaneously broken. The next step, $SU(2)_R\otimes U(1)_{B-L} \to U(1)_Y$, is achieved by $\langle \eta_R \rangle \propto v_R= 15$ TeV. At last, the SM group is broken by the two neutral components of $\Phi$ that acquire vevs satisfying the relation $(v_1^2+v_2^2)^{1/2} =v_{EW} = 246$ GeV. Step by step the Higgs mechanism is at play and, in the end, six would-be Goldstone bosons are absorbed by the gauge sector. The anomalous nature of the $U(1)_{PQ}$ symmetry, on the other hand, gives rise to a pseudo-Goldstone boson instead, the axion $a$, upon its spontaneous breaking by $v_\sigma$. Furthermore, the strong CP problem is solved with the PQ mechanism, and the axion with mass given by eq. (\ref{maxion}) becomes a good cold dark matter candidate. Among the remainder eleven scalar degrees of freedom, there exist a very heavy CP-even field, $S_\sigma$, with mass proportional to $v_\sigma$, and $S_1$, a $125$ GeV Higgs boson, identified as the SM one. All the other massive scalars, including two charged fields, $H^\pm_1$ and $H^\pm_2$, get masses proportional to the $v_R$ scale and can in principle be produced at colliders.

In the gauge sector of our LR model, we have twice as many massive gauge bosons as in the SM, as expected. With $v_R = 15$ TeV, the new neutral and charged gauge bosons, with masses of $5.8$ TeV and $4.9$ TeV, respectively, are heavy enough to evade the current experimental limits.

In the fermion sector, we have shown that since the $U(1)_{PQ}$ symmetry distinguishes $\Phi$ from $\tilde{\Phi}$, the corresponding tree-level CKM matrix presents no mixing and therefore does not agree with what the experimental picture tells us. This issue has been solved by considering non-renormalisable Yukawa interactions suppressed by the largest mass scale in our model, $\Lambda$. With the new terms in the Yukawa sector proportional to the suppression factor $\epsilon = v_\sigma/(\sqrt{2}\Lambda) \ll 1$, the resulting quark mixing has been shown to be naturally small. By assuming a particular point in the parameter space, given by eq. (\ref{v1v2e}) together with the Appendix \ref{AppB}, we have shown that our model can accommodate the current data on quark masses and mixing. As for the leptons, the charged lepton masses come mainly from renormalisable operators, while the neutrino masses depend directly on $\sigma$ via higher-order operators. The inverse seesaw mechanism is implemented, so that in addition to the three sub-eV active neutrinos, six TeV scale Majorana neutrinos, forming three quasi-Dirac pairs, are present. With the realisation of the inverse seesaw in our TeV scale model, non-standard $0\nu2\beta$ contributions, involving the new neutrinos and charged gauge boson, can become sizeable and need to be checked. Considering the benchmark solution in Appendix \ref{AppA}, we have calculated the most relevant $0\nu2\beta$ contributions and found that the standard $0\nu2\beta$ contribution, $\mathcal{A}_\nu$, is by far the dominant one. Coming from a solution with normal neutrino mass ordering, this contribution is, however, too small to be observed by the upcoming $0\nu2\beta$ experiments. New contributions to lepton flavour violating processes, such as leptonic decays $l_i\to l_j \gamma$, also arise and can be large in our framework. For the benchmark considered, the branching ratio for $\mu\to e \gamma$ is found to be just below the current experimental limit in a region that will soon be experimentally accessible.

\section*{Acknowledgements}
The work of A. G. D. was supported by the Conselho Nacional de Desenvolvimento Cient\'{\i}fico e Tecnol\'ogico (CNPq), grant No. 306636/2016-6.
J. L. acknowledges financial support under grant 2017/23027-2, S\~ao Paulo Research Foundation (FAPESP). The authors would like to thank Celso Nishi for useful comments and suggestions.

\appendix\label{Appendix}

\section{Scalar mass matrices}\label{AppA}

We show here the squared mass matrices of the scalar sector. Upon diagonalisation, we find the mass states and eigenvalues presented in section \ref{sec.3}.

Among the four charged fields in the scalar sector, three mix. In the basis $(\eta_L^\pm\,, \phi_1^\pm\,, \phi_2^\pm)$, we can write the following squared mass matrix
\be 
M^2_{\pm} =\frac{1}{2} (\lambda_{\eta \Phi 2}-\lambda_{\eta \Phi 1})
\begin{pmatrix}
v_1^2-v_2^2 & -v_2 v_R & -v_1 v_R \\
-v_2 v_R & \frac{v_2^2 v_R^2}{v_1^2-v_2^2} & \frac{v_1^2 v_R^2}{v_1^2-v_2^2} \\
-v_1 v_R & \frac{v_1 v_2 v_R^2}{v_1^2 - v_2^2} &\frac{v_1^2 v_R^2}{v_1^2 - v_2^2}
\end{pmatrix}
\ee 

Considering the CP-odd fields, in the basis $(A_\sigma\,, A_{1}\,, A_{2})$, we have
\be 
M^2_{A} =\frac{v_R^2(\lambda_{\eta \Phi 2}-\lambda_{\eta \Phi 1}) + 2(v_1^2-v_2^2) (\lambda_{\Phi345}-\lambda_{\Phi1}) }{2 (v_1^2-v_2^2)} 
\begin{pmatrix}
\frac{v_1^2v_2^2}{v_\sigma^2} & -\frac{v_1 v_2^2}{v_\sigma} & -\frac{v_1^2 v_2}{v_\sigma} \\
-\frac{v_1 v_2^2}{v_\sigma} & v_2^2  & v_1 v_2 \\
-\frac{v_1^2 v_2}{v_\sigma} & v_1 v_2 & v_1^2
\end{pmatrix}~,
\ee 
with $\lambda_{\Phi345}= \lambda_{\Phi3}+\lambda_{\Phi4}+\lambda_{\Phi5}$.
Finally, in the basis $(S_\sigma\,,S_1\,,S_2\,,S_R)$, the CP-even squared mass matrix is
\be 
M^2_{S} =\frac{1}{2 (v_1^2-v_2^2)} 
\begin{pmatrix} s_{11} & s_{12} & s_{13} & s_{14}\\
s_{12} & s_{22}  & s_{23} & s_{24} \\
s_{13} & s_{23} & s_{33} & s_{34} \\
s_{14} & s_{24} & s_{34}  & s_{44}
\end{pmatrix}~,
\ee 
with
\bea
s_{11}&=&
\frac{(v_1^2-v_2^2)[4 \lambda_\sigma v_\sigma^4  + 2 (\lambda_{\Phi345}-\lambda_{\Phi1}) v_1^2 v_2^2] + v_R^2v_1^2 v_2^2(\lambda_{\eta\Phi2}-\lambda_{\eta\Phi1})}{v^2_\sigma}~,\\
s_{12}&=&2 \lambda_{\sigma\Phi}v_1 v_\sigma (v_1^2-v_2^2)+2 (\lambda_{\Phi1}-\lambda_{\Phi345})\frac{v_1 v_2^2 (v_1^2-v_2^2)}{v_\sigma} + (\lambda_{\eta\Phi1}-\lambda_{\eta\Phi2})\frac{v_1 v_2^2 v_R^2}{v_\sigma} ~,\nn\\
s_{13}&=&2 \lambda_{\sigma\Phi}v_2 v_\sigma (v_1^2-v_2^2)+2 (\lambda_{\Phi1}-\lambda_{\Phi345})\frac{ v_1^2v_2 (v_1^2-v_2^2)}{v_\sigma} + (\lambda_{\eta\Phi1}-\lambda_{\eta\Phi2})\frac{v_1^2 v_2 v_R^2}{v_\sigma} ~,\nn\\
s_{14}&=&2 \lambda_{\sigma \eta} (v_1^2 - v_2^2) v_R v_\sigma ~,\nn\\
s_{22}&=&4(\lambda_{\Phi1}+\lambda_{\Phi2})v_1^4+2(\lambda_{\Phi1}-\lambda_{\Phi345})v_2^4+(\lambda_{\eta\Phi2}-\lambda_{\eta\Phi1})v_2^2v_R^2\nn\\
&&+ 2(\lambda_{\Phi345}-3\lambda_{\Phi1}-2\lambda_{\Phi2})v_1^2 v_2^2~,\nn\\
s_{23}&=&~(\lambda_{\eta \Phi1}-\lambda_{\eta \Phi2})v_1 v_2 v_R^2 + 2 (\lambda_{\Phi1}+2\lambda_{\Phi2}+\lambda_{\Phi345}) v_1 v_2 (v_1^2-v_2^2)~,\nn\\
s_{24}&=&~2 (\lambda_{\eta\Phi}+\lambda_{\eta\Phi1}) (v_1^2 - v_2^2) v_1 v_R ,\nn\\
s_{33}&=&-4(\lambda_{\Phi1}+\lambda_{\Phi2})v_2^4-2(\lambda_{\Phi1}-\lambda_{\Phi345})v_1^4+(\lambda_{\eta\Phi2}-\lambda_{\eta\Phi1})v_1^2v_R^2\nn\\
&&-2(\lambda_{\Phi345}-3\lambda_{\Phi1}-2\lambda_{\Phi2})v_1^2 v_2^2~~,\nn\\
s_{34}&=&2 (\lambda_{\eta\Phi}+\lambda_{\eta\Phi2}) (v_1^2 - v_2^2) v_2 v_R~,\nn\\
s_{44}&=&4 \lambda (v_1^2-v_2^2)v_R^2 ~.\nn
\eea

\section{Quark sector benchmarks}\label{AppB}

When the Yukawa matrices in eq. (\ref{qmass}) are given by 
\begin{eqnarray}
{\bf y}^q &=& \begin{pmatrix}
-0.01021 - 0.00522\, i & -0.01397 + 0.07163\, i & -0.01538 + 
  0.10359\, i \\ -0.01397 + 0.07163\, i & 0.3169 & 
 0.4565 + 0.0033\, i\\-0.01538 + 0.10359\, i & 0.4565 + 0.0033\, i & 
 0.6631 + 0.0094\, i\end{pmatrix}~,\\
{\bf h}^q &=& \begin{pmatrix}
1.2429 - 0.6707\, i & -1.635 + 0.250\, i & -1.842 + 0.048\, i \\
-1.635 + 0.250\, i & 1.0758 - 0.0427\, i & 0.0391 - 0.2405\, i \\ 
-1.842 + 0.048\, i &  0.0391 - 0.2405 \,i & -1.820 - 0.610\,i 
\end{pmatrix}\nn~,
\end{eqnarray}
the quark matrices ${\bf M}_u$, ${\bf M}_d$ can be diagonalised using 
\bea
|{\bf V}_{CKM}| &=& \begin{pmatrix}
 0.97446 & 0.22453 & 0.00364 \\
 0.22438 & 0.97359 & 0.04214 \\
 0.00896 & 0.04134 & 0.99910
\end{pmatrix}~~\mbox{and}~~J_{CP} =3.17\times10^{-5}~, \\
{\bf D}_L &=& \begin{pmatrix}
  0.2855 & 0.9480 & 0.0573 - 0.1288\, i\\
  0.7866 + 0.0217\, i & -0.2019 + 
  0.0721 \,i & -0.5787\\
  -0.5463 + 0.0301\, i & 0.2130 + 0.1001\,i & -0.8033
\end{pmatrix}~,\nn\\
{\bf U}_L &=& \begin{pmatrix}
  0.4915 & 0.8613 - 0.0054\, i & 
 0.02041 - 0.1270 \,i \\ 0.7204 + 0.0354 \, i & -0.3975 + 
  0.0654\, i & -0.5634 - 0.0003\, i\\ -0.4855 + 0.0491 \, i & 
 0.2961 + 0.0906 \, i & -0.8161 - 0.0056 i
\end{pmatrix}~,\nn
\eea 
satisfying the relation ${\bf V}_{CKM} = {\bf U}_L^\dagger {\bf D}_L $.

After diagonalisation the up-type and down-quark mass matrices become
\bea
{\bf m}_u &=& \mbox{diag}(m_u,\, m_c,\, m_t) = \mbox{diag} (2.2\times 10^{-3},\, 1.275,\, 173)~\mbox{GeV} \\
{\bf m}_d &=& \mbox{diag}(m_d,\,m_s,\,m_b) = \mbox{diag}(4.7\times 10^{-3},\, 95\times10^{-3},\, 4.18)~\mbox{GeV}.\nn
\eea

\section{Lepton sector benchmark}\label{AppC}

The active neutrino mass matrix in eq. (\ref{mnun}) is diagonalised by the PMNS matrix (${\bf U}$) which can be parametrised in terms of three mixing angles: $\theta_{12}$, $\theta_{13}$ and $\theta_{23}$, and three phases $\delta$, $\alpha_{21}$ and $\alpha_{31}$. Below, we present a benchmark solution using the current best-fit values for the ${\bf U}$ parameters as given in ref. \cite{Tanabashi:2018oca}.

This solution displays of a normal mass ordering and is obtained when the Yukawa matrices in eq. (\ref{mnun}) assume the following values
    \begin{eqnarray}
{\bf h} &=&\mbox{diag}(1.688,\,  0.01747,\,  0.00254)~,\\
{\bf y} &=& \begin{pmatrix}
0.05331 - 0.00734\, i &  -0.00152 - 0.00584 \,i &  -0.00114 + 
  0.00061\, i \\
  0.02933 + 0.08659\, i &  -0.0763 - 0.2305\, i &  -0.07405 + 
  0.05448\, i \\
  -0.02549 + 0.02928\, i &  
 0.04857 + 0.09974\, i &  -1.1143 - 0.8226\, i
\end{pmatrix}\nn~,
\end{eqnarray}
so that the active neutrino mass matrix eq. (\ref{mnuISS}) is diagonalised by the ${\bf U}$ mixing matrix defined in terms of its three mixing angles, a Dirac phase, for which we assume the best fit values as in ref. \cite{Tanabashi:2018oca}
\be
\sin^2{\theta_{12}} = 0.297 ~,~~\sin^2{\theta_{23}} =0.425~,~~\sin^2{\theta_{13}} =0.0215~,~~\delta = 1.38 \,\pi~,
\ee
and two Majorana phases, which for this specific solution are given by
\be
\alpha_{21} = 3.088~,~~\alpha_{31} = 0.270~,
\ee
giving rise to the following physical masses 
\be
{\bf m}_\nu =\mbox{diag} (0.0090,\, 0.01244,\, 0.05139)~\mbox{eV}~.
\ee 
Therefore, the sum of the masses and the effective Majorana mass are
\bea 
\sum_{i=1}^3 m_i = 0.07284~\mbox{eV}~~\mbox{and}~~\langle m_{ee} \rangle = 2.14 \times 10^{-3}~\mbox{eV}~.
\eea 
Whereas the masses of the heavy quasi-Dirac neutrino pairs are
\be
m^h_i= (0.532,\, 2.750,\,14.78)~~~\mbox{TeV}~,
\ee
with small splitting of order $\mu$.

The remaining $3\times 3$ mixing matrices in eq. (\ref{U2}) are
\bea 
{\bf V_{11}} &=& \begin{pmatrix} 
0.1263 - 0.6935\, i & 0.6935 + 0.1263\, i & 
  0.00761 + 0.05569 \,i \\0.04485 + 0.03378\, i & -0.03378 + 0.04485\, i & 
  0.6656 + 0.2269\, i \\-0.00132 + 0.00244\, i & -0.00244 - 
   0.00132\, i & -0.02026 + 0.04428\, i
\end{pmatrix}\,; \\
{\bf V_{12}} &=& \begin{pmatrix}
0.05569 - 0.00761 \, i & -0.00137 - 0.00007 \, i & -0.00007 + 
   0.00137 \, i \\0.2269 - 0.6656 \, i & -0.04194 + 0.02486 \, i & 
  0.02486 + 0.04194 \, i \\0.04428 + 0.02026 \, i & 0.2977 + 0.6395 \, i & 
  0.6395 - 0.2977 \, i
\end{pmatrix}\,; \\ 
{\bf V_{21}} &=& \begin{pmatrix}
 0.6601 \, i &
  0.6601 & -0.0004 - 0.2525 \, i \\-0.0037 + 0.2514 \, i & 
  0.2514 + 0.0037 \, i & -0.0138 + 0.6583 \, i \\-0.02795 + 0.01627 \, i &
  0.01627 + 0.02795 \, i & -0.01038 + 0.05119 \, i
\end{pmatrix}\,; \\
{\bf V_{22}} &=& \begin{pmatrix} 
0.2525 - 0.0004 \, i & -0.02137 + 0.00752 \, i & -0.00752 - 
   0.02137 \, i \\-0.6583 - 0.0138 \, i & -0.02900 - 0.04918 \, i &
  0.04918 - 0.02900 \, i \\-0.05119 - 0.01038 \, i &
  0.1408 + 0.6902 \, i & -0.6902 + 0.1408 \, i
\end{pmatrix}\,. 
\eea
The suppressed matrix ${\bf a} = {\bf m_D} ({\bf m_R}^\dagger)^{-1}$ present in the neutrino mixing matrix in Eq. (\ref{U1}) is
\be
{\bf a} \simeq \begin{pmatrix} 
0.036 & 0.013 & 
- 0.002\, i \\-0.18 + 0.048\, i & -0.64 - 
  1.55 \, i & -0.1 - 0.082\,i\\ 0.042 +
  0.082 \, i & -1.47 + 1.16 \, i & -4.05 - 2.81 \, i 
\end{pmatrix} \times 10^{-3}~,
\ee
whereas ${\bf b} = \textbf{{\boldmath$\mu$}} ({\bf m_R}^*)^{-1}$ is, as expected, even more suppressed with entries varying from $10^{-6}$ to $10^{-10}$. Finally, we explicitly show the $3\times3$ matrices ${\bf aV_{21}}$ and ${\bf aV_{22}}$ dictating the mixing between the active neutrinos ($\nu_{iL}$) and the heavy ones ($N_{iL}$ and $S_{iL}$), as can be inferred from Eq. (\ref{CC}), 
\bea\label{lhmix}
{\bf aV_{21}} &\simeq&\begin{pmatrix} 
0.027 \, i & 0.027  & 
  -4\times10^{-4}\, i \\ 0.37 - 0.28 \, i & -0.28 - 0.37\, i & 
 1.05 - 0.36 \, i \\ -0.18 - 0.33 \, i & -0.33 + 
 0.18 \, i & -0.54 - 1.17\, i
\end{pmatrix} \times 10^{-3}~,\\
{\bf aV_{22}} &\simeq& \begin{pmatrix} 
4.2\times10^{-4} & -2\times10^{-6} &  
   -2\times10^{-6}\, i\\
 0.36 + 1.05\, i & -0.012 & 
 - 0.012\, i\,\\
 1.17 - 0.54 \, i & 1.5 - 3.2 \, i & 3.2 + 1.5\, i
\end{pmatrix} \times 10^{-3}~.\nn
\eea

\section{Branching ratio for the process $l_i\to l_j \gamma$}\label{AppD}

We present here the branching ratios associated with the lepton flavour violating decays: $l_i\to l_j \gamma$. We adapt the expressions given in Ref.~\cite{Barry:2013xxa} to our case, and consider, for simplicity, that the contributions mediated by gauge fields are dominant over those mediated by the scalar fields. The latter fields can be taken to be heavier than $W_R$ without loss of generality. The general expression is given by
\be
BR(l_i \to l_j \gamma) = \frac{3 \alpha_{em}}{2 \pi}\left(|\left(G_L^\gamma\right)_{ij}|^2+|\left(G_R^\gamma\right)_{ij}|^2\right)~,
\ee
with
\bea\label{GLR}
\left(G_L^\gamma\right)_{ij} &=& \sum_{k=1}^{3}\left\{\left({\bf V_{11}}\right)_{ik}\,\left({\bf V_{11}}\right)^*_{jk} \left[ |\zeta|^2 \,G_1^\gamma\left(\frac{M_{N_k}^2}{m_{W_L}^2}\right)+ \frac{m_{W_L}^2}{m_{W_R}^2}\,G_1^\gamma\left(\frac{M_{N_k}^2}{m_{W_R}^2}\right)\right] \right.\\
&&\left. +\left({\bf V_{12}}\right)_{ik}\, \left({\bf V_{12}}\right)^*_{jk} \left[ |\zeta|^2\, G_1^\gamma\left(\frac{M_{S_k}^2}{m_{W_L}^2}\right)+ \frac{m_{W_L}^2}{m_{W_R}^2}\,G_1^\gamma\left(\frac{M_{S_k}^2}{m_{W_R}^2}\right)\right] \right.\nn\\
&& \left.+ \left({\bf a V_{21}}\right)^*_{ik}\,\left({\bf V_{11}}\right)^*_{jk}\,\frac{\zeta\,M_{N_k}}{m_\mu}\, G_2^\gamma\left(\frac{M_{N_k}^2}{m_{W_L}^2}\right)\right.\nn\\
&& \left. + \left({\bf a V_{22}}\right)^*_{ik}\,\left({\bf V_{12}}\right)^*_{jk}\,\frac{\zeta\,M_{S_k}}{m_\mu}\,   G_2^\gamma\left(\frac{M_{S_k}^2}{m_{W_L}^2}\right)\right\}~,\nn\\
\left(G_R^\gamma\right)_{ij} &=& \sum_{k=1}^{3}\left[ \left({\bf aV_{21}}\right)^*_{ik}\,\left({\bf aV_{21}}\right)_{jk}\, G_1^\gamma\left(\frac{M_{N_k}^2}{m_{W_L}^2}\right) + \left({\bf aV_{22}}\right)^*_{ik}\,\left({\bf aV_{22}}\right)_{jk}\, G_1^\gamma\left(\frac{M_{S_k}^2}{m_{W_L}^2}\right)\right.\nn\\
&& \left.+  \left({\bf V_{11}}\right)_{ik}\,\left({\bf a V_{21}}\right)_{jk}\,\frac{\zeta\, M_{N_k}}{m_\mu}\,  G_2^\gamma\left(\frac{M_{N_k}^2}{m_{W_L}^2}\right) \right.\nn\\
&& \left. + \left({\bf V_{12}}\right)_{ik}\,\left({\bf a V_{22}}\right)_{jk}\,\frac{\zeta\,M_{S_k}}{m_\mu}\, G_2^\gamma\left(\frac{M_{S_k}^2}{m_{W_L}^2}\right) \right]~,\nn
\eea 
where $G_1^\gamma(x)$ and $G_2^\gamma(x)$ are the following loop functions
\bea
G_1^\gamma(x) &=& -\frac{2 x^3 + 5 x^2 - x}{4(1-x)^3}-\frac{3 x^3}{2(1-x)^4}\log(x)~,\\
G_1^\gamma(x) &=& \frac{x^2 + -11 x + 4}{2(1-x)^2}-\frac{3 x^2}{(1-x)^3}\log(x)~.\nn
\eea 
\bibliographystyle{JHEP}
\bibliography{references}

\end{document}